\documentclass{iopart}
\pdfoutput=1
\usepackage{graphicx,rotating,subfigure,delarray,setstack,iopams}

\def\12{\frac{1}{2}}

\eqnobysec

\makeatother
\begin{document}
\bibliographystyle{iopart-num}
\title{Boundary Fidelity and Entanglement in the symmetry protected topological phase of the SSH model}

\author{J. Sirker} \address{Department of Physics and Research Center
  OPTIMAS, TU Kaiserslautern, D-67663 Kaiserslautern, Germany;
  Department of Physics and Astronomy, University of Manitoba,
  Winnipeg, Manitoba R3T 2N2, Canada}
\author{M.  Maiti} 
\address{Department of Physics and Research Center OPTIMAS, TU
  Kaiserslautern, D-67663 Kaiserslautern, Germany; Bogoliubov
  Laboratory of Theoretical Physics, Joint Institute for Nuclear
  Research, 141980 Dubna, Moscow Region, Russia}
\author{N. P.  Konstantinidis}
\address{Department of Physics and Research Center OPTIMAS, TU
  Kaiserslautern, D-67663 Kaiserslautern, Germany; Max-Planck-Institut
  f\"ur Physik komplexer Systeme, 01187 Dresden, Germany}
\author{N. Sedlmayr} \address{Department of Physics and Research
  Center OPTIMAS, TU Kaiserslautern, D-67663 Kaiserslautern, Germany;
  Institute de Physique Th\'eorique, CEA/Saclay, Orme des Merisiers,
  91190 Gif-sur-Yvette Cedex, France} \date{\today}

\begin{abstract}
  We present a detailed study of the fidelity, the entanglement
  entropy, and the entanglement spectrum, for a dimerized chain of
  spinless fermions---a simplified Su-Schrieffer-Heeger (SSH)
  model---with open boundary conditions which is a well-known example
  for a model supporting a symmetry protected topological (SPT) phase.
  In the non-interacting case the Hamiltonian matrix is tridiagonal
  and the eigenvalues and -vectors can be given explicitly as a
  function of a single parameter which is known analytically for odd
  chain lengths and can be determined numerically in the even length
  case. From a scaling analysis of these data for essentially
  semi-infinite chains we obtain the fidelity susceptibility and show
  that it contains a boundary contribution which is different in the
  topologically ordered than in the topologically trivial phase.  For
  the entanglement spectrum and entropy we confirm predictions from
  massive field theory for a block in the middle of an infinite chain
  but also consider blocks containing the edge of the chain.  For the
  latter case we show that in the SPT phase additional
  entanglement---as compared to the trivial phase---is present which
  is localized at the boundary. Finally, we extend our study to the
  dimerized chain with a nearest-neighbour interaction using exact
  diagonalization, Arnoldi, and density-matrix renormalization group
  methods and show that a phase transition into a topologically
  trivial charge-density wave phase occurs.
\end{abstract}


\maketitle

\tableofcontents

\section{Introduction}
The Su-Schrieffer-Heeger (SSH) model has been introduced as a
tight-binding model to describe conducting polymers such as
polyacetylene \cite{SSH,SuSchriefferHeegerRMP}. A simplified version
of the SSH Hamiltonian for spinless fermions and a static lattice
dimerization is given by
\begin{equation}
\label{SSH}
\!\!\!\!\!\!\!\!\!\!\!\!\!\!\!\!\!\!\!\!\!\!\!\!\!\!\!\! H=-t\sum_{j=1}^{N-1} [1+(-1)^j\delta](c_{j+1}^\dagger c_j + \mbox{h.c.})+U\sum_{j=1}^{N-1}(n_j-1/2)(n_{j+1}-1/2)
\end{equation}
where $t$ is the hopping amplitude, $\delta$ the dimerization
parameter, $U$ the nearest-neighbour repulsion, and $N$ the number of
lattice sites. The creation operator for a spinless fermion at site
$j$ is denoted by $c_j^\dagger$, and $n_j=c_j^\dagger c_j$ is the
occupation number operator. The model possesses time reversal symmetry
$T$ and particle-hole symmetry $C$ with $T^2=1$ and $C^2=1$ and thus,
for $U=0$, belongs into the BDI class in a classification scheme of
single-particle Hamiltonians \cite{RyuSchnyder}.

We will concentrate first on the non-interacting case, $U=0$, and
postpone a discussion of the interacting case to
section~\ref{Interacting}. For half-filling and any finite $\delta$
(dimerized case) the excitation spectrum is then gapped with a gap
$\Delta_E\propto |\delta|$ in the thermodynamic limit. It has also
been realized early on that the excitations in the dimerized case are
topological solitons and anti-solitons (domain walls between the two
possible dimerization patterns) which carry exactly half an electron
charge \cite{SuSchriefferHeegerRMP,JackiwRebbi}.

In recent years, the SSH model has also attracted interest as one of
the simplest examples for a model with a symmetry protected
topological (SPT) phase \cite{RyuHatsugai02,Wen12}. The topologically
non-trivial phase for the Hamiltonian (\ref{SSH}) with $N$ even is
realized for $\delta>0$. It can only be transformed into a topological
trivial phase by either breaking the symmetries which protect it or by
closing the excitation gap. Thus gapless edge modes have to be present
at the boundary of an SSH chain with $\delta>0$ and the topological
trivial vacuum. An open SSH chain with $\delta<0$, on the other hand,
is topologically trivial and there are no edge modes. Instead of
distinguishing the two phases by the presence or absence of edge
modes, one can equivalently consider the Zak-Berry phase $\gamma$
(bulk-boundary correspondence) which is a $\mathbb{Z}_2$ bulk
invariant with $\gamma=\pi$ in the SPT phase and $\gamma=0$ in the
topologically trivial phase
\cite{Berry,Zak,RyuHatsugai06,ViyuelaRivas}.

The complexity of the ground state of a quantum many-body system and
its response to changes in the microscopic parameters of the
Hamiltonian can be characterized by its entanglement properties and by
the fidelity, respectively. An interesting question is the relation
between these quantum information measures and SPT order. For model
(\ref{SSH}) without interactions the von-Neumann entanglement entropy
$S_{\rm ent}$ as a function of dimerization $\delta$ in a periodic
chain, obtained after tracing out one half of the system, has been
investigated in Ref.~\cite{RyuHatsugai06}. A detailed study of the
entanglement entropy and spectrum as a function of the block size in
the SPT and topological trivial phases has, however, not been
performed yet. While it is clear that the ground state in the SPT
phase is still only short-ranged entangled so that the bulk
entanglement properties are the same as in the trivial phase
\footnote{Note, however, that tracing out part of the system creates
  virtual edges so that the entanglement spectra will have different
  degeneracies depending on whether a strong or a weak bond is cut.
  The degeneracies of the entanglement spectrum can be used to
  classify topological phases in one-dimensional interacting systems
  \cite{FidkowskiKitaev,PollmannTurner,TurnerPollmann}.}, additional
entanglement can arise at the boundaries. Studying these boundary
contributions to the entanglement is one of the main goals of our
study.

The fidelity susceptibility $\chi$ near phase transitions has also
been studied in various models with topological phases
\cite{AbastoHamma,ShuoGu}, however, these studies concentrated on the
bulk contribution $\chi_0$. For an open system there will in addition
also be a {\it boundary contribution}, $\chi=\chi_0+\chi_B/N$, where
$N$ is the number of lattice sites, which to the best of our knowledge
has not been investigated for models with SPT phases before. It is
exactly this boundary contribution $\chi_B$ which should be sensitive
to the presence or absence of edge states.

Our paper is organized as follows: In section~\ref{Model} we review
the exact solution of the SSH chain for periodic boundary conditions
(PBC) and open boundary conditions (OBC). In section~\ref{Fidelity} we
calculate the fidelity susceptibility and analyze the boundary
contribution $\chi_B$ both in the SPT and in the topologically trivial
phase. In section~\ref{Entanglement_entropy} we first briefly review
known results for the entanglement entropy in critical and gapped
quantum chains and the general formalism to calculate the entanglement
spectrum for free fermion models. In the analysis of the obtained data
for the SSH model we then show, in particular, that predictions from
massive field theory for the scaling of $S_{\rm ent}$ with block size
are not valid in the SPT phase. In section~\ref{Entanglement_spectra}
we study the entanglement spectra and compare with known analytical
results in the limit of large block sizes. By using exact
diagonalization, Arnoldi, and density-matrix renormalization group
methods we then extend our investigations in section~\ref{Interacting}
to the SSH model with a nearest-neighbour Coulomb repulsion. On the
basis of the entanglement properties we identify, in particular, a
transition into a topologically trivial charge-density wave phase. In
section~\ref{Conclusions} we summarize our main results and conclude.

\section{Diagonalization of the SSH model}
\label{Model}
We will briefly review the steps required to diagonalize the SSH model
with periodic and open boundary conditions and give the eigenenergies
and eigenvectors. For OBC we will consider both the case where the
number of sites $N$ is even and the case where it is odd.  The
calculations to obtain the fidelity susceptibility and the
entanglement spectrum and entropy are then explained in
section~\ref{Fidelity} and
sections~\ref{Entanglement_entropy}-\ref{Entanglement_spectra},
respectively.
\subsection{Periodic boundary conditions and $N$ even}
\label{PBC_diag}
This is the simplest case to handle. Taking the doubling of the unit
cell due to the dimerization into account and performing a Fourier
transform, the Hamiltonian (\ref{SSH}) can be written as
\begin{equation}
\label{PBC1}
\!\!\!\!\!\!\!\!\!\!\!\!\!\!\! H=-2t\sum_{k}\left(\begin{array}{c}c_{1k}^\dagger \\ c_{2k}^\dagger\end{array}\right)^T
 \left(\begin{array}{cc}0 & \cos k+ i\delta\sin k\\ \cos k - i\delta\sin k & 0\end{array}\right)
 \left(\begin{array}{c}c_{1k} \\ c_{2k}\end{array}\right)
\end{equation}
where $k=2\pi n/N$ and $n=1,\cdots,N/2$. This is straightforwardly
diagonalized. The eigenvalues are
 \begin{equation}
 \lambda^\pm_k=\pm 2t\sqrt{\cos^2 k+\delta^2\sin^2 k }=\pm \sqrt{2}t\sqrt{1+\delta^2+(1-\delta^2)\cos 2k}
 \end{equation}
 with a gap at the Fermi points $k_F=\pm\pi/2$ given by $\Delta_E =
 4t|\delta|$. The new operators in which the Hamiltonian is diagonal, 
\begin{equation}
\label{PBC2}
H=-\sum_k \lambda^+_k \left(\alpha_k^\dagger\alpha_k -\beta_k^\dagger\beta_k\right)\, ,
\end{equation}
are given by
\begin{equation}
\alpha_k=\frac{1}{\sqrt{2}} (\mathcal{A} c_{k1} +c_{k2})\quad ,\quad \beta_k=\frac{1}{\sqrt{2}} (-\mathcal{A} c_{k1} +c_{k2})
\end{equation}
with $\mathcal{A}=\lambda_k^+/[2t(\cos k+i\delta\sin k)]$. The ground
state of the Hamiltonian (\ref{PBC2}) at half-filling is obtained by
filling up the valence band, $|\Psi_0\rangle = \prod_k
\alpha_k^\dagger |0\rangle$.

\subsection{Open boundary conditions and $N$ odd}
\label{OBC_odd}
While in the PBC case only two momentum modes are coupled so that we
are left with the diagonalization of a $2\times 2$ matrix, see
equation~(\ref{PBC1}), open boundary conditions instead lead to a
coupling of all momentum modes so that a Fourier transform does not
diagonalize the problem.

In real space, on the other hand, the Hamiltonian is a symmetric
tridiagonal matrix with zero diagonal elements and superdiagonal and
subdiagonal elements which alternate between $1\pm \delta$. The
eigenvalues and -vectors are explicitly known if the number of lattice
sites is odd \cite{Shin}. The eigenvalues are
\begin{equation}
\label{EVs}
\lambda_0 = 0 \quad , \quad \lambda^\pm_k = \pm \sqrt{2}t\sqrt{1+\delta^2+(1-\delta^2)\cos \theta_k}
 \end{equation}
 where $\theta_k=2\pi k/(N+1)$ and $k=1,\cdots,(N-1)/2$. The spectrum
 thus contains a single zero energy mode. If we parametrize each eigenvector as 
\begin{equation}
|\psi_k\rangle=(x_1^k,x_2^k,\ldots,x_N^k)^{\rm{T}}\,,
\end{equation}
then the not-normalized elements of the eigenvector for the zero
energy mode are
\begin{equation}
\label{N_odd_zero}
\!\!\!\!\!\!\!\!\!\!\!\!\!\!\!\!\!\!\!\!\!\!\!\!\!\!\!\!\!\!\!\!\! x^0_{2n-1}=\left(-\frac{1-\delta}{1+\delta}\right)^{n-1}, \; n=1,\cdots,\frac{N+1}{2};\quad x^0_{2n}=0,\; n=1,\cdots,\frac{N-1}{2}.
\end{equation}
For the other modes one finds
\begin{eqnarray}
\label{N_odd_rest}
x^k_{2n-1} &=& \frac{1+\delta}{1-\delta}\sin\left(\frac{2\pi k(n-1)}{N+1}\right)+\sin\left(\frac{2\pi k n}{N+1}\right), \; n=1,\cdots, \frac{N+1}{2}\nonumber \\
x^k_{2n} &=& \frac{\lambda_k^\pm}{1-\delta}\sin\left(\frac{2\pi kn}{N+1}\right),\; n=1,\cdots,\frac{N-1}{2}. 
\end{eqnarray}
Depending on the sign of the dimerization $\delta$ the edge mode is thus localized either at the right or the left end of the chain.

\subsection{Open boundary conditions and $N$ even}
\label{OBC_even}
For an even number of lattice sites we expect two edge states for
$\delta>0$ and no edge states for $\delta<0$. The eigenenergies
$\lambda_k^\pm$ can still be written as in equation~(\ref{EVs}), however, the
parameters $\theta_k$ are no longer known analytically and instead have
to be determined by numerically solving the implicit equation \cite{Shin}
\begin{equation}
(1-\delta)T(\theta_k,N/2)+(1+\delta)T(\theta_k,N/2-1)=0 \label{thetaeqn}\,,
\end{equation}
with
\begin{equation}
\label{implicit}
T(\theta_k,n)=\frac{\sin[(n+1)\theta_k]}{\sin\theta_k}\,.
\end{equation}
The eigenvector corresponding to each of the $N$ solutions $\theta_k$ of
equation~(\ref{thetaeqn}) is then given by
\begin{eqnarray}
\label{states_Neven}
x_{2n-1}^k(\delta)&=&\frac{1+\delta}{1-\delta}T(\theta_k,n-2)+T(\theta_k,n-1)\nonumber \\
x_{2n}^k(\delta)&=&\frac{\lambda^\pm_k}{1-\delta}T(\theta_k,n-1)\,,
\end{eqnarray}
where $n$ takes values from $1,2,\cdots N/2$.

\begin{figure}
\begin{center}
\includegraphics*[width=0.75\columnwidth]{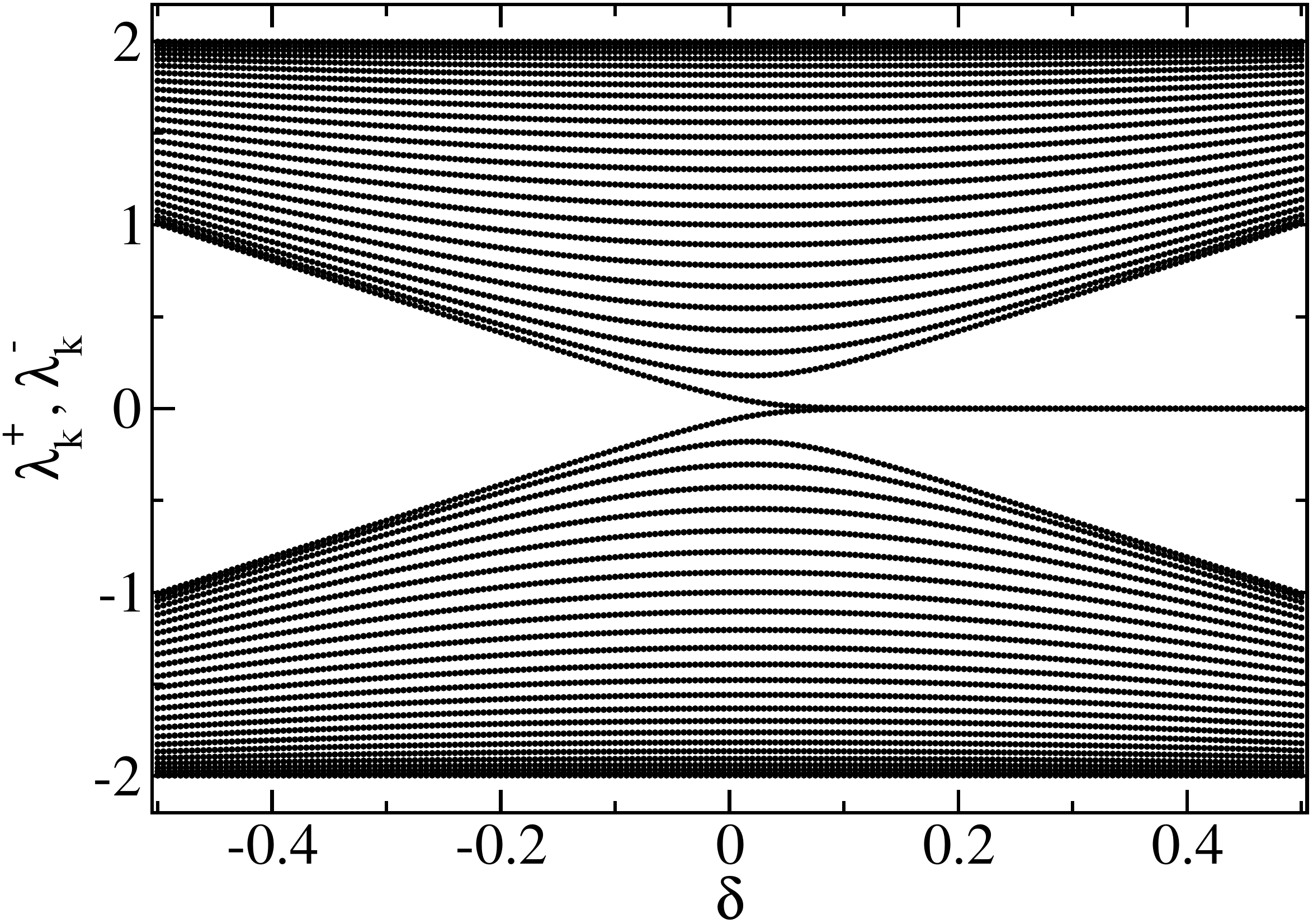}
\end{center}
\caption{Single-particle spectrum $\lambda_k^\pm$, equation
  (\ref{EVs}) with $\theta_k$ determined by equation (\ref{thetaeqn}),
  of an open SSH chain as a function of dimerization $\delta$ for
  $N=50$ sites.}
\label{Spectrum50}
\end{figure}

While for $\delta<0$ all $N$ states are extended, localized edge
states are possible for $\delta>0$ corresponding to complex solutions
of equation~(\ref{thetaeqn}). More precisely, we find that the system has
$N-2$ real solutions and two complex solutions if $\delta>\delta_c$ where 
\begin{equation}
 \delta_c=\frac{1}{1+N}, \label{deltac}
\end{equation}
while all solutions are real if $\delta<\delta_c$. Physically, this
can be understood as follows: The extended states in a finite system
are protected against a small dimerization by the finite size gap
$\Delta_E$ of order $1/N$. Only if the gap $\Delta_E\sim |\delta|$
induced by the dimerization overcomes the finite size gap can the
spectrum change qualitatively. Alternatively, one can also think about
this problem in terms of the localization length $\xi_{\rm loc}\sim
1/\delta$ of the edge state, see equation~(\ref{N_odd_zero}), which
has to become smaller than the system size in order to allow for
localized states.  In the thermodynamic limit, $N\to\infty$, we have
$\delta_c\to 0$ so that the phase transition between the topological
trivial and the SPT phase occurs, as expected, at $\delta=0$. The
eigenspectrum for a finite system as a function of dimerization shown
in figure~\ref{Spectrum50} makes it clear that there is always a small
gap between the two edge modes which will only go to zero for all
$\delta>0$ in the thermodynamic limit.

\section{Bulk and boundary fidelity susceptibility}
\label{Fidelity}
At a quantum phase transition between different states of matter we
expect the many-body wave function to change. A measure for this
change is the fidelity
$F(\delta,\epsilon)=|\langle\Psi(\delta)|\Psi(\delta+\epsilon)\rangle|$
where $\epsilon$ parameterizes the difference in dimerization between
the two states with $F(\delta,0)\equiv 1$. Similar to the Anderson
orthogonality catastrophe \cite{AndersonOC} we expect the overlap
between the two many-body wave functions to vanish exponentially with
system size $N$ for $\epsilon\neq 0$. It is therefore appropriate to
consider the fidelity density defined as
\begin{equation}
\label{fidelity}
f(\delta,\epsilon)=-\frac{1}{N}\ln F(\delta,\epsilon)\, .
\end{equation}
Since $f(\delta,0)\equiv 0$ is a minimum, the fidelity has the small
$\epsilon$ expansion $f(\delta,\epsilon)=\chi(\delta)\epsilon^2
+\mathcal{O}(\epsilon^3)$ where 
\begin{equation}
\label{fidelity_susci}
\chi(\delta)=\frac{1}{2}\frac{\partial^2 f}{\partial\epsilon^2}\bigg|_{\epsilon=0}
\end{equation}
is the fidelity susceptibility. At a quantum phase transition, this
quantity shows universal scaling behaviour
\cite{ZanardiPaunkovic,VenutiZanardi,RamsDamski,SirkerFidelity} and
can thus be used to characterize the transition.

Here we want to use the fidelity susceptibility to characterize the
transition from the topologically trivial into the SPT phase in the
SSH model (\ref{SSH}). In an expansion in $1/N$ we can write
$\chi=\chi_0 +\chi_B/N +\mathcal{O}(N^{-2})$. While the bulk
contribution $\chi_0=\lim_{N\to\infty} \chi$ will be the same in both
phases, we expect that the {\it boundary fidelity susceptibility}
$\chi_B$ is different.

For a system of non-interacting electrons, the many-body wave functions
are simple Slater determinants of the single particle eigenstates. The
fidelity $F(\delta,\epsilon)$ can thus be expressed as
\begin{equation}
\label{Fidelity_overlap}
F(\delta,\epsilon)=|\det A_{kl}(\delta,\epsilon)|\,,
\end{equation}
with the $M\times M$ single particle overlap matrix
\begin{equation}
\label{Overlap_matrix}
A_{kl}(\delta,\epsilon)=\langle\psi_k(\delta)|\psi_l(\delta+\epsilon)\rangle\,,
\end{equation}
where $M$ is the number of particles in the ground state. The fidelity
susceptibility can now be obtained as
\begin{equation}
\label{fidelity_susci2}
\chi(\delta)=-\frac{1}{N}\lim_{\epsilon\to 0}\frac{\ln F(\delta,\epsilon)}{\epsilon^2}
\end{equation}
and thus does not require taking numerical derivatives.

\subsection{Periodic boundary conditions}
In this case there is no boundary so that $\chi_B=0$. Furthermore, we
expect $\chi(\delta)=\chi(-\delta)$ for $N$ even because changing the
sign of $\delta$ is in this case equivalent to a simple relabelling of
the lattice sites.

Using the results of section \ref{PBC_diag}, the analytic calculation
of the fidelity at half-filling is straightforward
\begin{eqnarray}
|\langle \Psi(\delta)|\Psi(\delta+\epsilon)\rangle| &=& \prod_{k,k'} |\langle 0 | \alpha_k\tilde\alpha_{k'}^\dagger |0\rangle | \\
&=& \frac{1}{2}\prod_{k,k'} |\langle 0 | (\mathcal{A}c_{1k}+c_{2k})(\tilde\mathcal{A}^*c^\dagger_{1k'}+c^\dagger_{2k'}) |0\rangle | \nonumber \\
&=& \prod_k \frac{|\mathcal{A}\tilde\mathcal{A}^*+1|}{2} \nonumber
\end{eqnarray}
where $\tilde\alpha_k$ and $\tilde \mathcal{A}$ are obtained by
replacing $\delta\to\delta+\epsilon$. For the fidelity density this
leads to
\begin{equation}
f=-\frac{1}{N}\sum_k\ln\frac{|\mathcal{A}\tilde\mathcal{A}^*+1|}{2} \stackrel{N\to\infty}{\to} -\frac{1}{\pi}\int_0^{\pi/2}\ln\frac{|\mathcal{A}\tilde\mathcal{A}^*+1|}{2} dk \, .
\end{equation}
The bulk fidelity susceptibility can now be calculated in closed form
\begin{equation}
\label{chi_bulk}
\chi_0=\frac{1}{32\pi}\int_0^{\pi/2}dk\frac{\sin^2 2k}{(\cos^2 k+\delta^2\sin^2 k)^2} = \frac{1}{32|\delta|(1+|\delta|)^2}\, .
\end{equation}
The finite size corrections in the gapped phase are exponentially
small, $\chi(N)-\chi_0\sim N\exp(-N/\xi)$ for $N\gg 1$, where $\xi$ is
the correlation length, see equation (\ref{Corrlength}). At the
critical point $\delta=0$ the fidelity susceptibility thus diverges as
$\chi_0\sim 1/|\delta|$. This is consistent with the general scaling
theory at a critical point $\delta_c$ which predicts $\chi_0\sim
|\delta -\delta_c|^{d\nu -2}$ where $d$ is the spatial dimension and
$\nu$ the critical exponent related to the divergence of the
correlation length at the critical point. In the case considered here
we have $\delta_c=0$, $d=1$, and $\nu=1$.

\subsection{Open boundary conditions and $N$ odd}
In the open chain case, we expect a $1/N$ boundary contribution
$\chi_B$ (surface term) to the fidelity susceptibility. According to
the bulk-boundary correspondence, a symmetry protected topological
phase as indicated by a bulk invariant Zak-Berry phase is related to
the presence of edge modes. We thus expect that $\chi_B$ is a measure
which can distinguish between a topologically trivial and a
non-trivial phase.

We start by considering the case of an odd number of sites $N$ where
the single particle eigenstates are known exactly, see section
\ref{OBC_odd}. Note that in this case a single edge mode is present
both for $\delta<0$ and for $\delta>0$. In order to obtain both the
bulk and the boundary contribution, we use
equations~(\ref{Fidelity_overlap}), (\ref{Overlap_matrix}), and
(\ref{fidelity_susci2}) with the exact single particle eigenstates
given by equations~(\ref{N_odd_zero}) and (\ref{N_odd_rest}). We are
interested in the half-filled case, however, for $N$ odd we can only
put either $(N-1)/2$ or $(N+1)/2$ particles into the system. This
means that the zero energy edge mode is either empty or occupied.
Because the Hamiltonian is particle-hole symmetric, the fidelity is,
however, exactly the same in both cases.

\begin{figure}
\begin{center}
\includegraphics*[width=0.75\columnwidth]{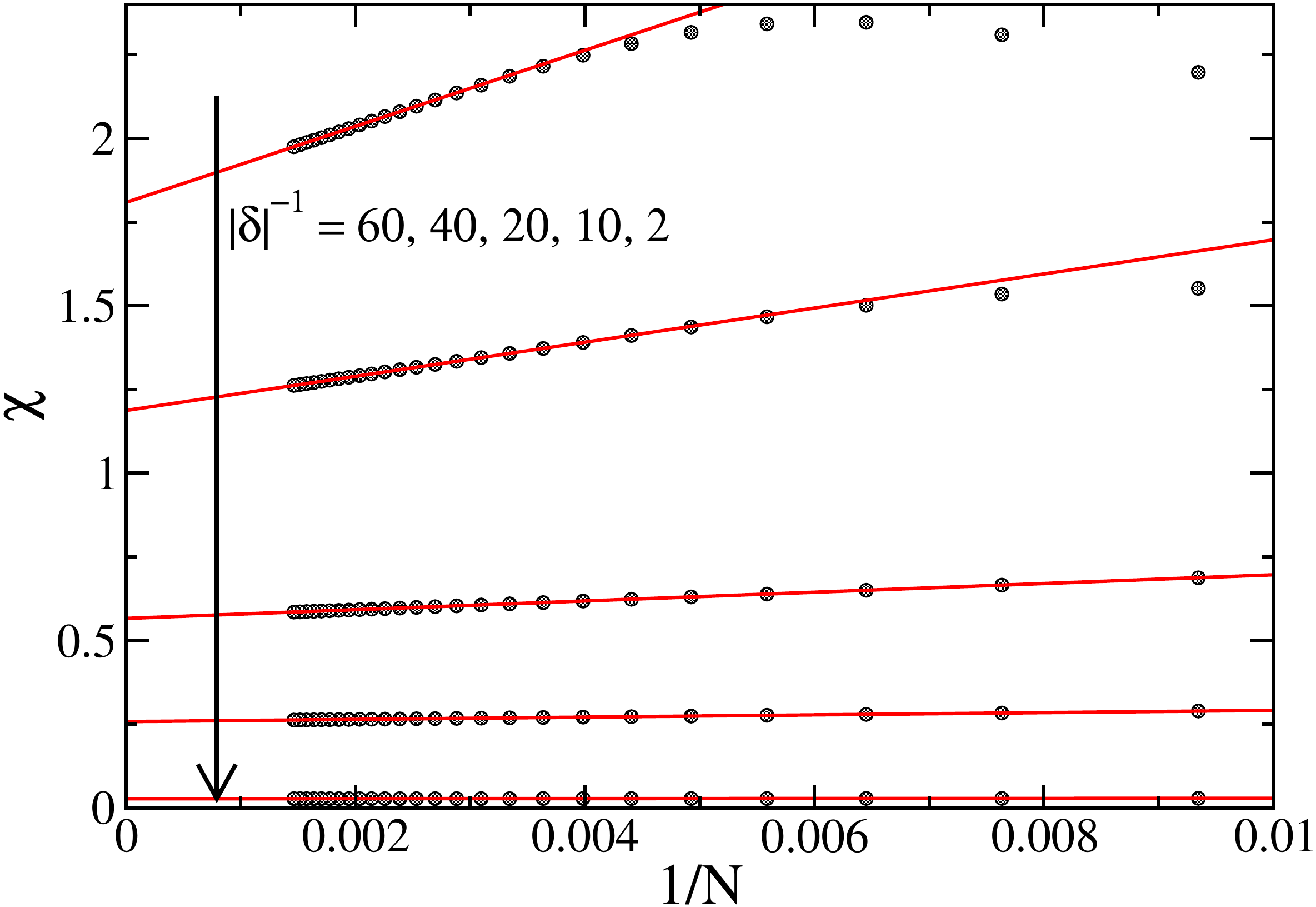}
\end{center}
\caption{Fidelity susceptibility for the SSH chain with OBC and an odd
  number of sites (symbols). The lines are linear fits for $1/N$
  small. Note that the next-leading corrections become more important
  with decreasing $|\delta|$ with the expansion in $1/N$ breaking down
  at the critical point $\delta=0$.}
\label{Fig_Nodd_fidelity_scaling}
\end{figure}

\begin{figure}
\begin{center}
\includegraphics*[width=0.75\columnwidth]{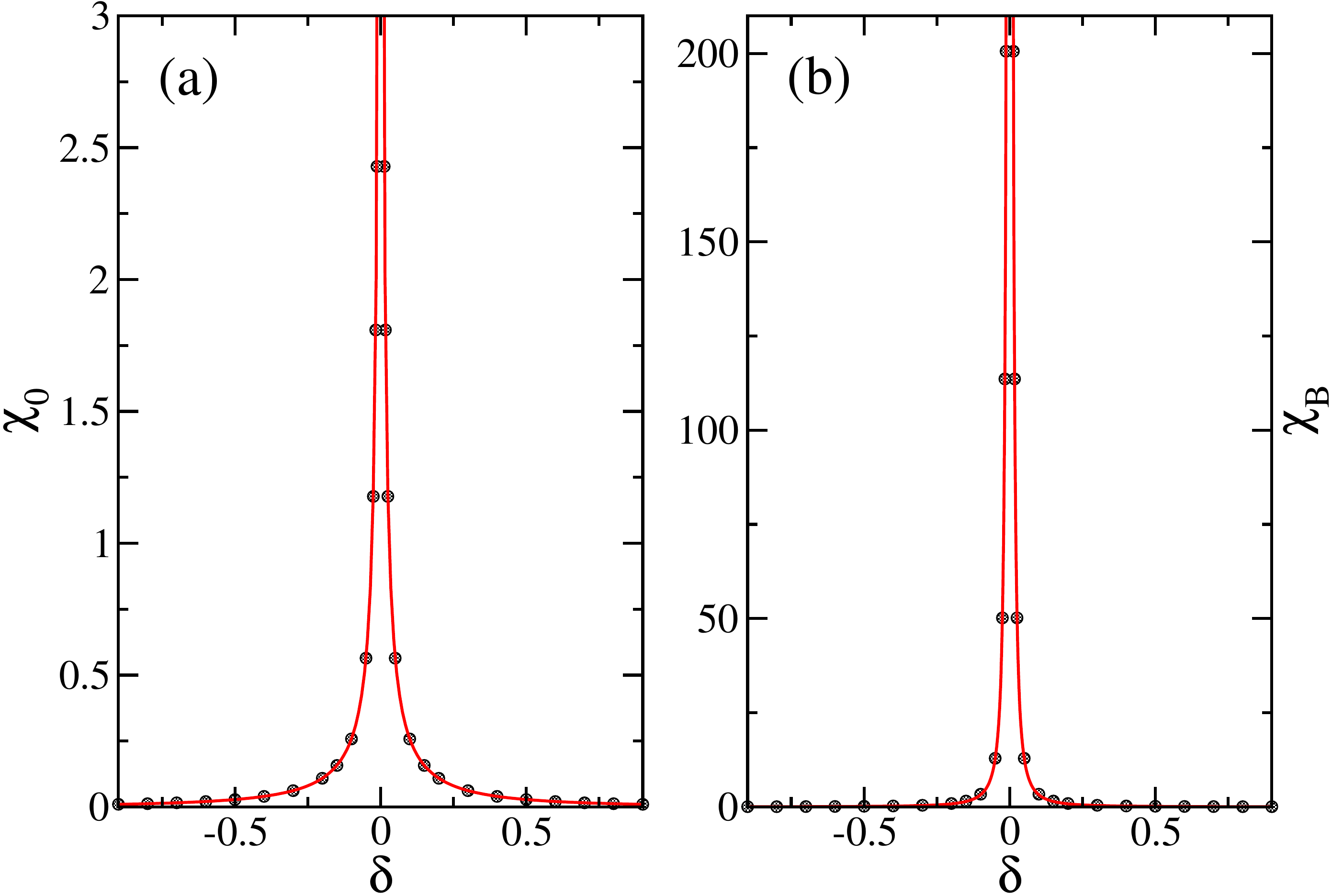}
\end{center}
\caption{Symbols denote (a) the bulk susceptibility, and (b) the
  boundary susceptibility for $N$ odd extracted from the scaling in
  figure~\ref{Fig_Nodd_fidelity_scaling}. The data in (a) are compared
  with the analytical solution (lines), equation~(\ref{chi_bulk}). In
  (b) the lines represent $\chi_B\approx 0.033 |\delta|^{-1.988}$
  obtained by fitting the data for $|\delta|\leq 0.2$.}
\label{Fig_Nodd_suscis}
\end{figure}

In figure~\ref{Fig_Nodd_fidelity_scaling} we show exemplarily the
scaling of the fidelity susceptibility for different dimerizations,
demonstrating that the leading correction to the bulk susceptibility
is indeed of order $1/N$.
The bulk and boundary susceptibilities extracted from the scaling are
shown in figure~\ref{Fig_Nodd_suscis}.
Note that both are symmetric in $\delta$ as they should be because
$\delta\to -\delta$ is---up to a relabelling of the lattice sites---a
symmetry of the Hamiltonian (\ref{SSH}) if $N$ is odd. The bulk
contribution $\chi_0$ is independent of the boundary conditions and
agrees with the analytical solution (\ref{chi_bulk}) obtained using
periodic boundary conditions. For the boundary contribution $\chi_B$
we numerically find a behaviour close to the critical point which is
consistent with $\chi_B\sim 1/|\delta|^\eta$ with $\eta\approx 2$.
Note, however, that we have two different edges for the $N$ odd case
which can give boundary contributions which scale differently. The
analysis of $\chi_B$ is therefore easier in the $N$ even case
considered next where either both edges have a localized state or both
are trivial.

\subsection{Open boundary conditions and $N$ even}
In the $N$ even case we can again calculate the overlap matrix
(\ref{Overlap_matrix}) from the single particle eigenstates,
equation~(\ref{states_Neven}). However, this time the parameters $\theta_k$
have to be determined by the implicit equation (\ref{thetaeqn}).
Especially for the bound states where $\theta_k$ becomes imaginary
this requires a high accuracy making the calculations numerically
challenging for large system sizes. 

\begin{figure}
\begin{center}
\includegraphics*[width=0.75\columnwidth]{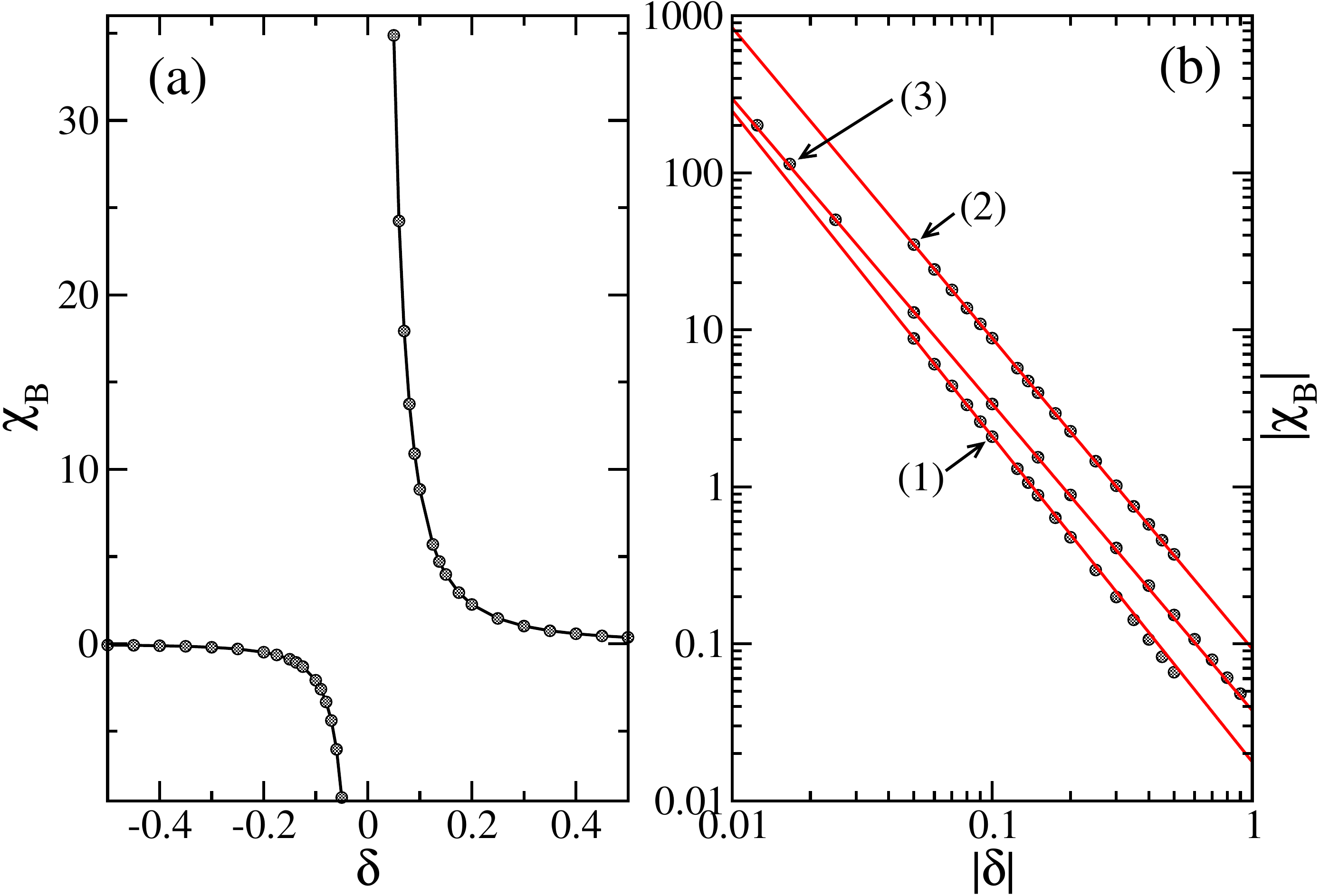}
\end{center}
\caption{(a) $\chi_B$ in the thermodynamic limit for an even number of
  sites (circles). Lines are a guide to the eye. (b) Fits of the data
  (circles) for (1) $N$ even, $\delta<0$: $|\chi_B^{(1)}|\approx
  0.0177/|\delta|^{2.073}$, (2) $N$ even, $\delta>0$:
  $\chi_B^{(2)}\approx 0.0928/\delta^{1.979}$, and (3) $N$ odd: data
  and function $\chi_B^{(3)}=(\chi_B^{(1)}+\chi_B^{(2)})/2$, see
  equation (\ref{edge_contributions}). For the fits the data with
  $|\delta|\leq 0.1$ have been used.}
\label{Fig_Neven_susci}
\end{figure}

The boundary susceptibility obtained in this case is shown in
figure~\ref{Fig_Neven_susci}.
Note that $\chi_B$ in the topological trivial phase ($\delta<0$) is
qualitatively different from $\chi_B$ in the SPT phase ($\delta>0$).
In particular, the sign is different. The data in
figure~\ref{Fig_Neven_susci}(b), furthermore, indicate that also the
exponents of, what seems to be a power-law divergence at the critical
point, are different by about 5\%. The presence or absence of edge
states is thus clearly visible in $\chi_B$.

For large chain lengths $N$ the edges should become independent so
that $\chi_B$ is simply the sum of the contributions from each of the
two edges. In particular, this implies that
\begin{equation}
\label{edge_contributions}
\!\!\!\!\!\!\!\!\!\!\lim_{N\to\infty}[\chi_B(N\;\mbox{even},\delta) + \chi_B(N\;\mbox{even},-\delta)]/2 = \chi_B(N-1\; \mbox{odd},\pm\delta).
\end{equation}
This behaviour is confirmed by the fits shown in figure
\ref{Fig_Neven_susci}(b). The data for the $N$ odd case are thus
better fitted by the sum of two power laws with different exponents
rather than by the single power law used in figure
\ref{Fig_Nodd_suscis}.

\begin{figure}
\begin{center}
\includegraphics*[width=0.75\columnwidth]{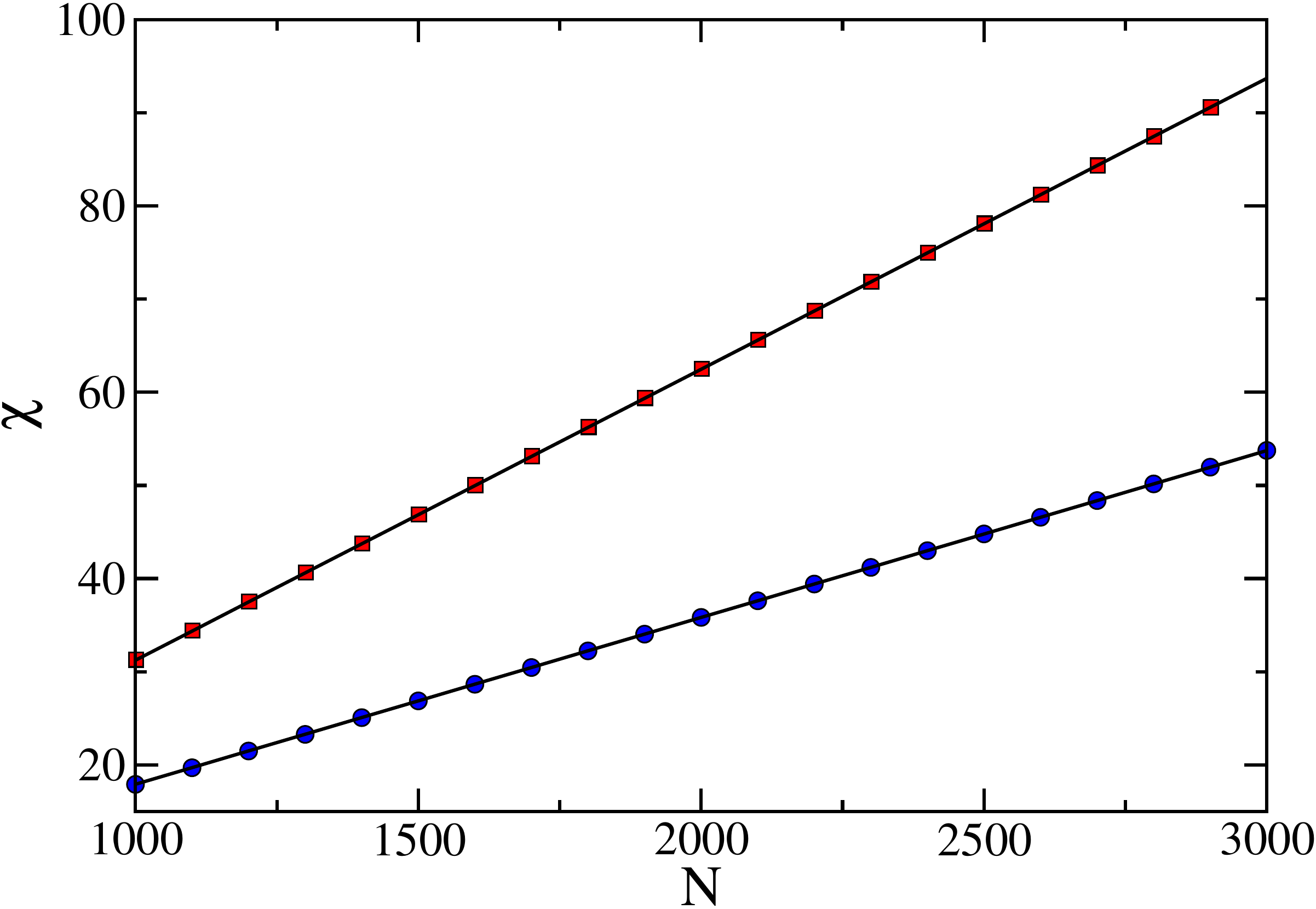}
\end{center}
\caption{Finite size scaling of $\chi$ at the critical point
  $\delta=0$ for $N$ even (circles) and $N$ odd (squares) with open boundary conditions.
  The lines are fits with $\chi\approx 0.01791 N$ ($N$ even), and
  $\chi\approx 0.03123 N\approx N/32$ ($N$ odd).}
\label{Fig_Scaling_at_delta00}
\end{figure}

Finally, we can also study the finite size scaling of the fidelity
susceptibility right at the critical point, see
figure~\ref{Fig_Scaling_at_delta00}.
While both data sets for $N$ even and $N$ odd show a scaling $\chi\sim
N$ consistent with the general scaling prediction $\chi\sim
N^{2/d\nu-1}$ \cite{RamsDamski}, the prefactors are different.

To summarize, we have established that the boundary susceptibility
$\chi_B$ behaves qualitatively different in the SPT ($\delta>0$) than
in the topologically trivial phase ($\delta<0$). In particular,
$\chi_B(\delta<0)<0$ and $\chi_B(\delta>0)>0$.

\section{Entanglement entropy}
\label{Entanglement_entropy}
The entanglement entropy is the von-Neumann entropy of a reduced density matrix
\begin{equation}
 S_A=-\Tr (\rho_A \log \rho_A)\,
\label{EE}
\end{equation}
with $\rho_A=\Tr_B \rho$ where $\rho=|\Psi\rangle\langle\Psi|$ is
the full density matrix with $|\Psi\rangle$ being the ground state
of the many-body Hamiltonian. The system has been divided up into two
blocks $A$ and $B$. It is easy to prove that $S_A=S_B\equiv
S_{\rm{ent}}$. The entanglement entropy is an effective measure to
characterize the ground states of many-body systems. A gain in
entanglement entropy can, furthermore, be a driving force for a
Peierls transition \cite{SirkerHerzog} leading to a dimerization of
the chain as considered here. For one-dimensional critical quantum
systems it has been shown that $S_{\rm{ent}}$ shows a universal
dependence on the length $\ell$ of the block $A$.  For a system of
size $N$ with PBC one finds
\begin{equation}
\label{EE2}
S^{CFT}_{\rm{ent}}=\frac{c}{3}\ln\left[\frac{N}{\pi}\sin\frac{\pi\ell}{N}\right]+s_1
\end{equation}
where $c$ is the central charge of the underlying conformal field
theory (CFT) in the scaling limit, and $s_1$ a non-universal constant
\cite{CalabreseCardy}. In the thermodynamic limit $N\to\infty$, the
CFT formula (\ref{EE2}) reduces to
\begin{equation}
\label{EE2prime}
S^{CFT}_{\rm{ent}}=\frac{c}{3}\ln\ell +s_1 \, .
\end{equation}
By a conformal mapping one can show that a similar relation also holds
for a system at small but finite temperatures
\cite{SirkerEntanglement}. For the undimerized chain of spinless
fermions, equation~(\ref{SSH}) with $\delta=0$, one finds $c=1$ and
$s_1\approx 0.726$ \cite{JinKorepin}.

These results are altered though if one changes the boundary
conditions from periodic to open. In this case the entanglement
entropy of a block which includes the end of the chain reads
\cite{CalabreseCardy}
\begin{equation}
\label{EE3}
S^{CFT}_{\rm{ent}}=\frac{c}{6}\ln\left[\frac{2N}{\pi}\sin\frac{\pi\ell}{N}\right]+\ln g+ s_1/2
\end{equation}
where $\ln g$ is the boundary entropy related to the ground state
degeneracy of the considered system \cite{AffleckLudwig}. For the
critical undimerized chain we have $g=1$. In numerical studies the
scaling law for OBC, equation~(\ref{EE3}), is, however, often obscured
by corrections coming from leading operators of the underlying CFT
\cite{LaflorencieSoerensen}. For critical systems described by
Luttinger liquid theory one finds, for example,
\begin{equation}
\label{EE4}
S_{\rm{ent}}\sim S^{CFT}_{\rm{ent}}+\cos(2k_F\ell)\ell^{-K}
\end{equation}
where $k_F$ is the Fermi momentum and $K$ the Luttinger liquid
parameter
\cite{CalabreseCampostrini,CardyCalabrese10,FagottiCalabrese,DalmonteErcolessi1,DalmonteErcolessi2}.

In a massive phase one expects, on the other hand, that the
entanglement entropy saturates once the block size become larger than
the correlation length $\xi$. If the system is in a massive phase but
close to a critical point then also the leading corrections can be
calculated and one finds for a block in an infinite chain
\begin{equation}
\label{EE5}
S_{\rm{ent}}=\frac{c}{3}\ln(\xi_1/a)+U-\frac{1}{8}\sum_{\alpha=1} K_0(2\ell/\xi_\alpha)
\end{equation}
where $U$ and $a$ are non-universal constants, $\xi_\alpha$ the
correlation lengths (with $\xi_1$ being the largest), and $K_0$ the
modified Bessel function \cite{CardyCastro-Alvaredo}. This type of
scaling behaviour has recently been confirmed numerically for certain
spin models \cite{LeviCastro-Alvaredo}.

\subsection{General setup}
In the following, we want to study the entanglement entropy of the SSH
model (\ref{SSH}). We will consider two cases: First, a block in an
infinite chain. Here we can simply use PBC which makes it possible to
perform the thermodynamic limit exactly. This case will provide a
numerical check of the formula (\ref{EE5}). Second, we will consider
the case of a block at the end of a semi-infinite chain.  Here we can
use either odd chain lengths where the eigenstates are known
explicitly, or even chain lengths. The odd chain length has an edge
state which is either filled or empty, which are equivalent by
particle-hole symmetry. In contrast at half-filling a finite open
system with an even number of sites in the SPT phase has two edge
states, only one of which is occupied at half filling. However, the
boundary states at the edges in the thermodynamic limit are a
superposition of states localized at the left and the right edge. Thus
we can also consider a semi-infinite chain with a partially occupied
edge state.  Since the correlation length $\xi\sim |\delta|^{-1}$ is
finite for $\delta\neq 0$ it suffices to consider chain lengths $N\gg
\xi$ in order to be effectively in the thermodynamic limit. We are
particularly interested in the difference in entanglement entropies
with and without a localized state being present at the boundary of
the chain.

For a free fermion model there is a general setup to obtain the
eigenvalues of the reduced density matrix $\rho_A$
\cite{ChungPeschel,Peschel02}. The main observation is that $\rho_A$
can be represented as the exponential of a free-fermion operator
$\mathcal{H}_A$---the so-called entanglement Hamiltonian---with
\begin{equation}
\label{rho_red}
\rho_A = \frac{1}{\mathcal{Z}}\exp(-\mathcal{H}_A)=\frac{1}{\mathcal{Z}}\exp\left(-\sum_{k=1}^\ell\epsilon_k a_k^\dagger a_k\right)
\end{equation}
and $\mathcal{Z}=\Tr\exp(-\mathcal{H}_A)$. On the other hand, all
properties of a free-fermion system can be obtained from the matrix
$\mathbf{C}$ of two-point correlation functions because all higher
correlation functions can be expressed through the two-point functions
by using Wick's theorem. The matrix elements of $\mathbf{C}$ for all
two-point functions within the considered block $A$ are, in
particular, determined by the reduced density matrix $\rho_A$, i.e.,
matrix elements are given by
\begin{equation}
\label{Cij}
C_{ij}=\Tr(\rho_A c_i^\dagger c_j) \, .
\end{equation}
The eigenvalues $\zeta_k$ of the correlation matrix can then be
related to the eigenvalues $\epsilon_k$ of the entanglement
Hamiltonian in equation (\ref{rho_red}) and one finds
\begin{equation}
\label{EV_relation}
\zeta_k = (\e^{\epsilon_k}+1)^{-1}.
\end{equation}
According to equation~(\ref{rho_red}) the $2^\ell$-many eigenvalues
$z_p$ of $\rho_A$ itself are now obtained by considering all possible
fillings of the $\ell$-many levels $\epsilon_k$
\begin{equation}
\label{EV_rho}
z_p=\frac{1}{\mathcal{Z}}\prod_{n^{(p)}}\exp\left(-\epsilon_k n^{(p)}_k\right) 
\end{equation}
where $n^{(p)}=(n^{(p)}_1,\cdots,n^{(p)}_\ell)$ is a vector of
single-level occupation numbers, $n^{(p)}_k\in \{0,1\}$, and
$\mathcal{Z}=\sum_p z_p$. The entanglement entropy is then given by
\begin{equation}
\label{Sent_EV}
S_{\rm{ent}}=-\sum_p z_p \ln z_p . 
\end{equation}

\subsection{A block in an infinite chain}
Using the results from section (\ref{PBC_diag}) for a chain with PBC
at half-filling, the two-point correlation function can be easily
calculated and the thermodynamic limit performed exactly. For even
distances one finds $C_{ij}=\delta_{ij}/2$, i.e., only the onsite
correlation function (occupation number) is nonzero. For odd distances
the result is
\begin{equation}
\label{Corr_PBC}
\!\!\!\!\!\!\!\!\!\!\!\!\!\!\!\!\!\!\!\!\!\!\!\!\!\!\!\!\!\!\!\!\!\!\!\!\!\!\!\!\!\!\langle c^\dagger_{2n}c_{2m+1}\rangle =\frac{1}{2\pi}\int_{-\pi/2}^{\pi/2} dk \frac{\cos k\cos k(2m+1-2n)+\delta\sin k\sin k(2m+1-2n)}{\sqrt{\cos^2k +\delta^2\sin^2 k}}
\end{equation}
and the correlation function $\langle c^\dagger_{2n+1}c_{2m}\rangle$
is obtained by replacing $\delta\to -\delta$. For the non-dimerized
case, this reduces to the well-known result
\begin{equation}
\label{Corr_usual}
\langle c^\dagger_{n}c_{m}\rangle =\frac{1}{\pi}\frac{\sin\left[\frac{\pi}{2}(m-n)\right]}{m-n} \, .
\end{equation}
By methods of steepest descent we can also obtain the asymptotic
behaviour of the correlation function (\ref{Corr_PBC}) at large
distances. In particular, we find that the dominant correlation length
is given by
\begin{equation}
\label{Corrlength}
\xi^{-1}\equiv \xi_1^{-1}=\frac{1}{2}\left|\ln\frac{1+\delta}{1-\delta}\right| \, .
\end{equation}

\begin{figure}
\begin{center}
\includegraphics*[width=0.75\columnwidth]{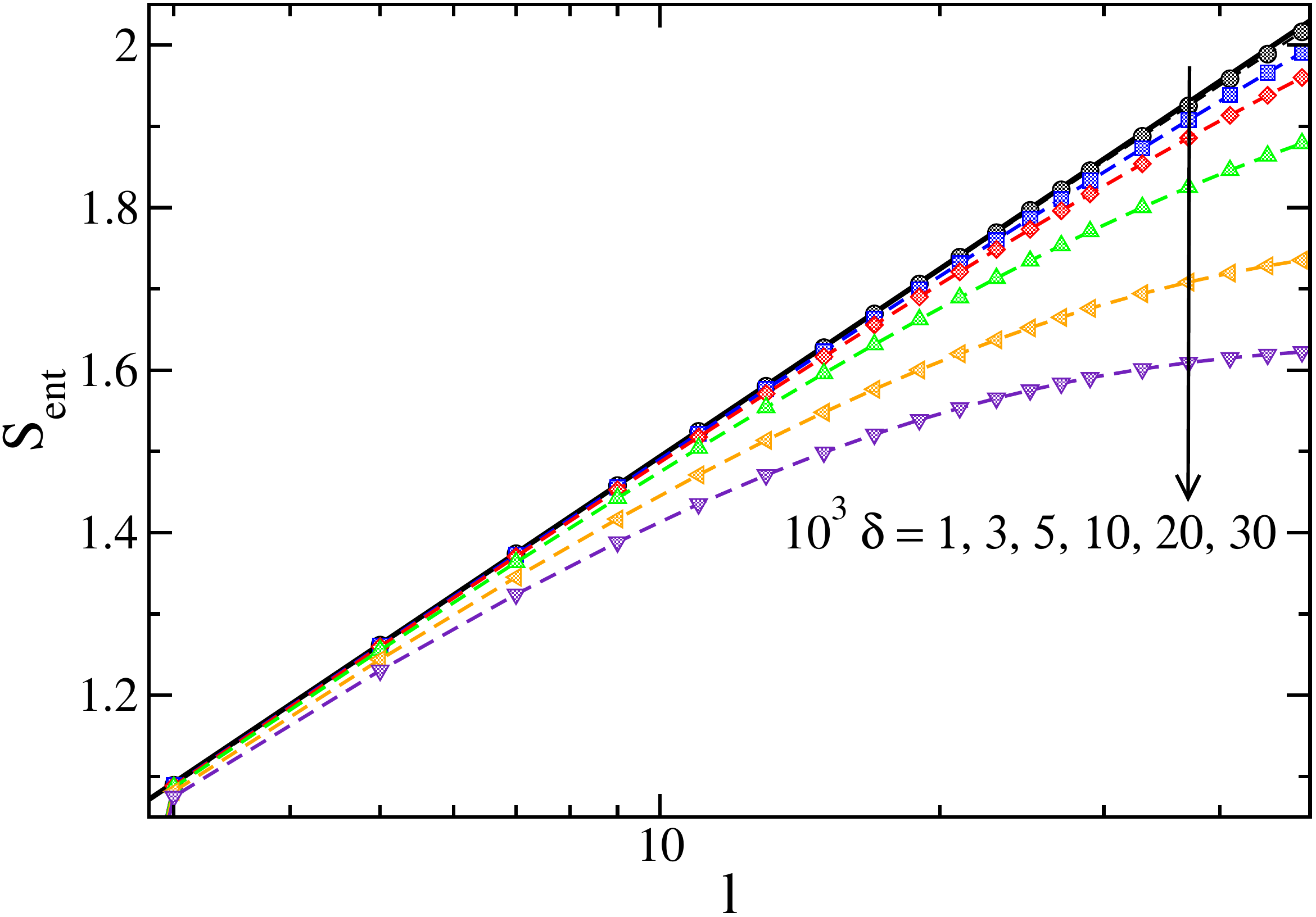}
\end{center}
\caption{$S_{\rm{ent}}$ for a block in an infinite chain and small
  dimerizations (symbols). The solid line is the CFT result,
  equation~(\ref{EE2}), using the value $s_1=0.726$ obtained for the
  non-dimerized chain \cite{JinKorepin}. The dashed lines are a guide
  to the eye.}
\label{Sent_small_dim}
\end{figure}

\begin{figure}
\begin{center}
\includegraphics*[width=0.75\columnwidth]{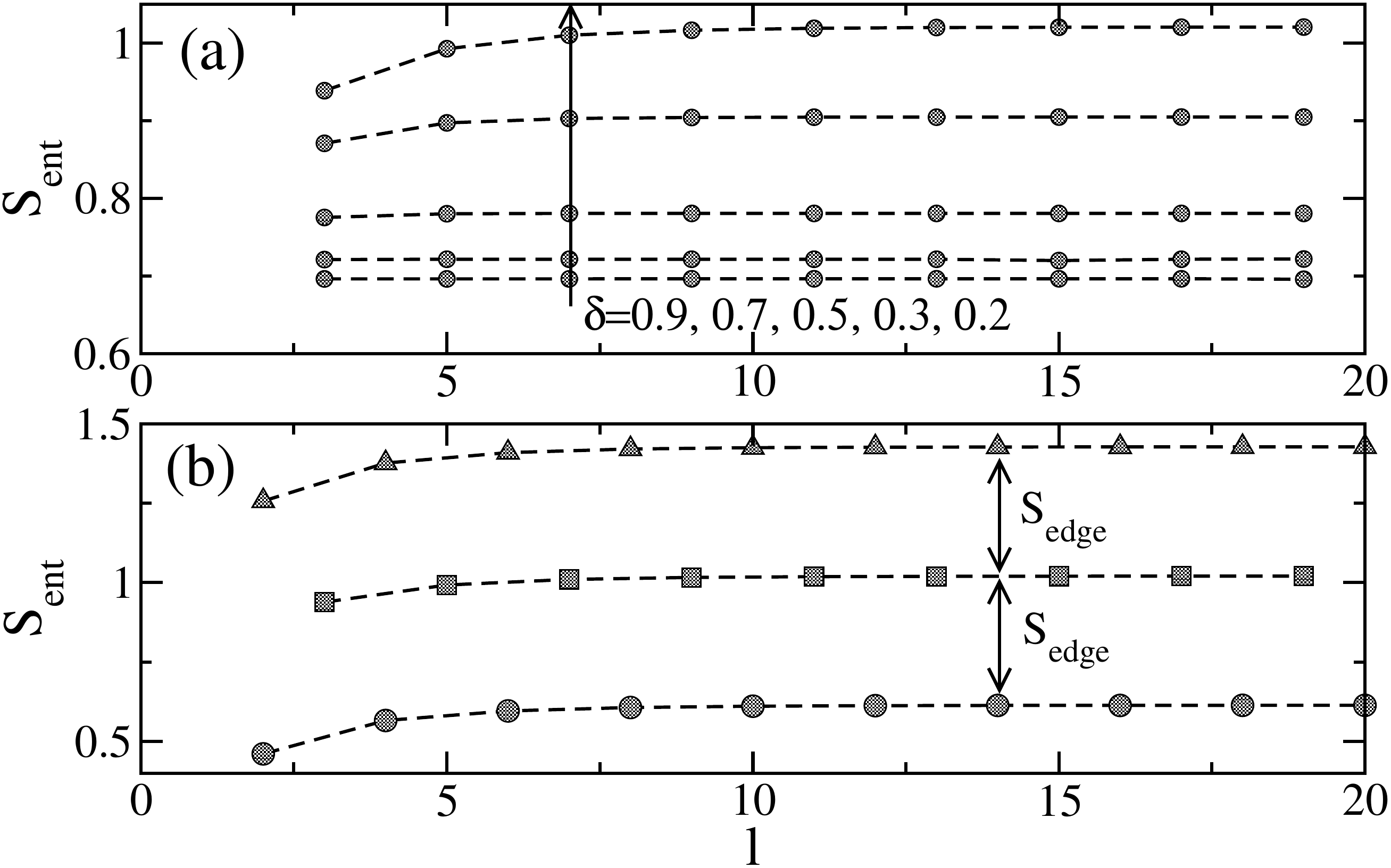}
\end{center}
\caption{(a) $S_{\rm{ent}}$ for a block with an odd number of sites in
  an infinite chain and large dimerizations. (b) $S_{\rm{ent}}$ for
  $\ell$ even and $\delta=-0.2$ (circles), $\ell$ odd (squares), and
  $\ell$ even and $\delta=+0.2$ (triangles). Each non-trivial edge
  gives a contribution $S_{\rm{edge}}$ to $S_{\rm{ent}}$. The dashed
  lines are a guide to the eye.}
\label{Sent_large_dim}
\end{figure}

\begin{figure}
\begin{center}
\includegraphics*[width=0.75\columnwidth]{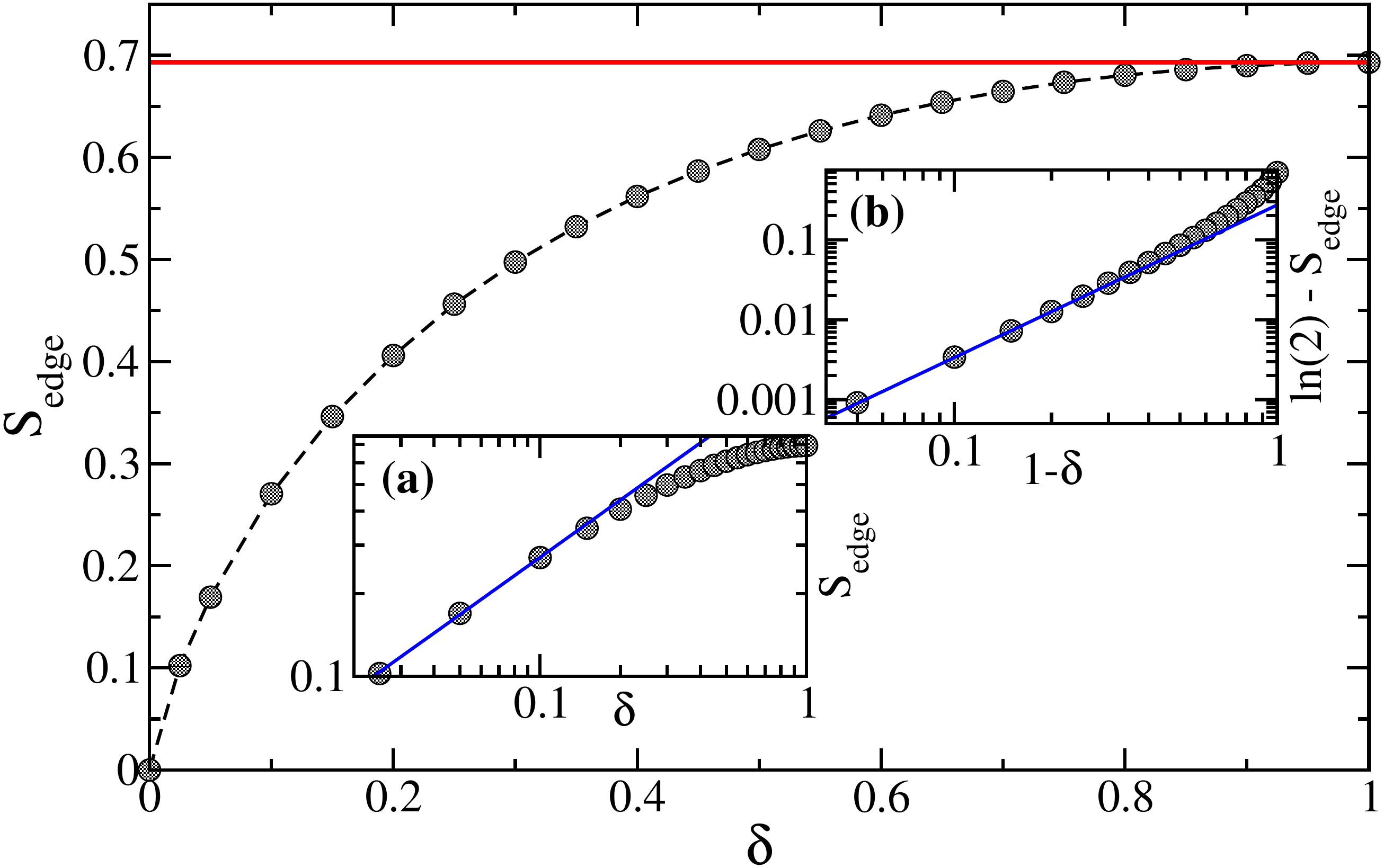}
\end{center}
\caption{Edge contribution (\ref{Sedge}) to the entanglement entropy
  of a block in an infinite chain. The line marks the limiting value
  $S_{\rm{edge}}^{\delta\to 1}=\ln 2$. The dashed line is a guide to
  the eye. Inset (a): Fit for $|\delta|\ll 1$ with $S_{\rm edge}\approx
  1.339|\delta|^{0.694}$. Inset (b): Fit for $|\delta|\lesssim 1$ with
  $S_{\rm edge}\approx \ln 2-0.273(1-|\delta|)^{1.909}$.}
\label{Fig_Sedge}
\end{figure}

There should now be three different regimes depending on the ratio of
block size $\ell$ and largest correlation length $\xi\equiv \xi_1$:
(1) For $\xi\gg \ell\gg 1$ the entanglement entropy will essentially
follow the scaling law for a critical system, equation~(\ref{EE2}), as
obtained by CFT. Exemplarily, we present in
figure~\ref{Sent_small_dim} results for $S_{\rm{ent}}$ for small
dimerizations and odd block lengths $\ell$. Indeed, we see that the
smaller the dimerization is the longer equation~(\ref{EE2})
approximately holds as a function of block size.
(2) For $\xi\ll \ell$, on the other hand, the entanglement entropy
will settle to a constant value $S_{\rm{ent}}=\frac{c}{3}\ln(\xi/a)+U$
where both constants $a$ and $U$ are non-universal. This behaviour is
shown exemplarily in figure~\ref{Sent_large_dim}(a).
In figure~\ref{Sent_large_dim}(b) we show, furthermore, that the
saturation value depends on whether two weak bonds ($\ell$ even), a
weak and a strong bond ($\ell$ odd), or two strong bonds ($\ell$ even)
are cut. We have chosen the block for $\ell$ even in such a way that
$\delta<0$ corresponds to cutting two weak and $\delta>0$ to cutting
two strong bonds. For block sizes $\ell\gg\xi$ each non-trivial
edge---caused by the cutting of a strong bond---gives an independent
contribution $S_{\rm{edge}}$ to the entanglement entropy. More
precisely, we can define the edge contribution as
\begin{equation}
\label{Sedge}
S_{\rm{edge}}=\lim_{\ell\to\infty}|S_{\rm{ent}}(\ell,\; \ell\;\rm{odd})-S_{\rm{ent}}(\ell+1,\; \ell+1\; \rm{even})|,
\end{equation}
which is shown as a function of dimerization in
figure~\ref{Fig_Sedge}.
In particular, in the limit $|\delta|\to 1$ the entanglement entropy
will be dominated entirely by the edge contribution so that
$S_{\rm{ent}}^{\delta\to 1}(\ell\; \rm{even})=2
S_{\rm{edge}}^{\delta\to 1}$, $S_{\rm{ent}}^{\delta\to 1}(\ell\;
\rm{odd})=S_{\rm{edge}}^{\delta\to 1}$, and $S_{\rm{ent}}^{\delta\to
  -1}(\ell\; \rm{even})\equiv0$ with $S_{\rm{edge}}^{\delta\to 1}=\ln
2$. The insets of figure~\ref{Fig_Sedge} are consistent with a power
law behaviour of $S_{\rm edge}$ for small $\delta$ and for $\delta\to
1$.

\begin{figure}
\begin{center}
\includegraphics*[width=0.75\columnwidth]{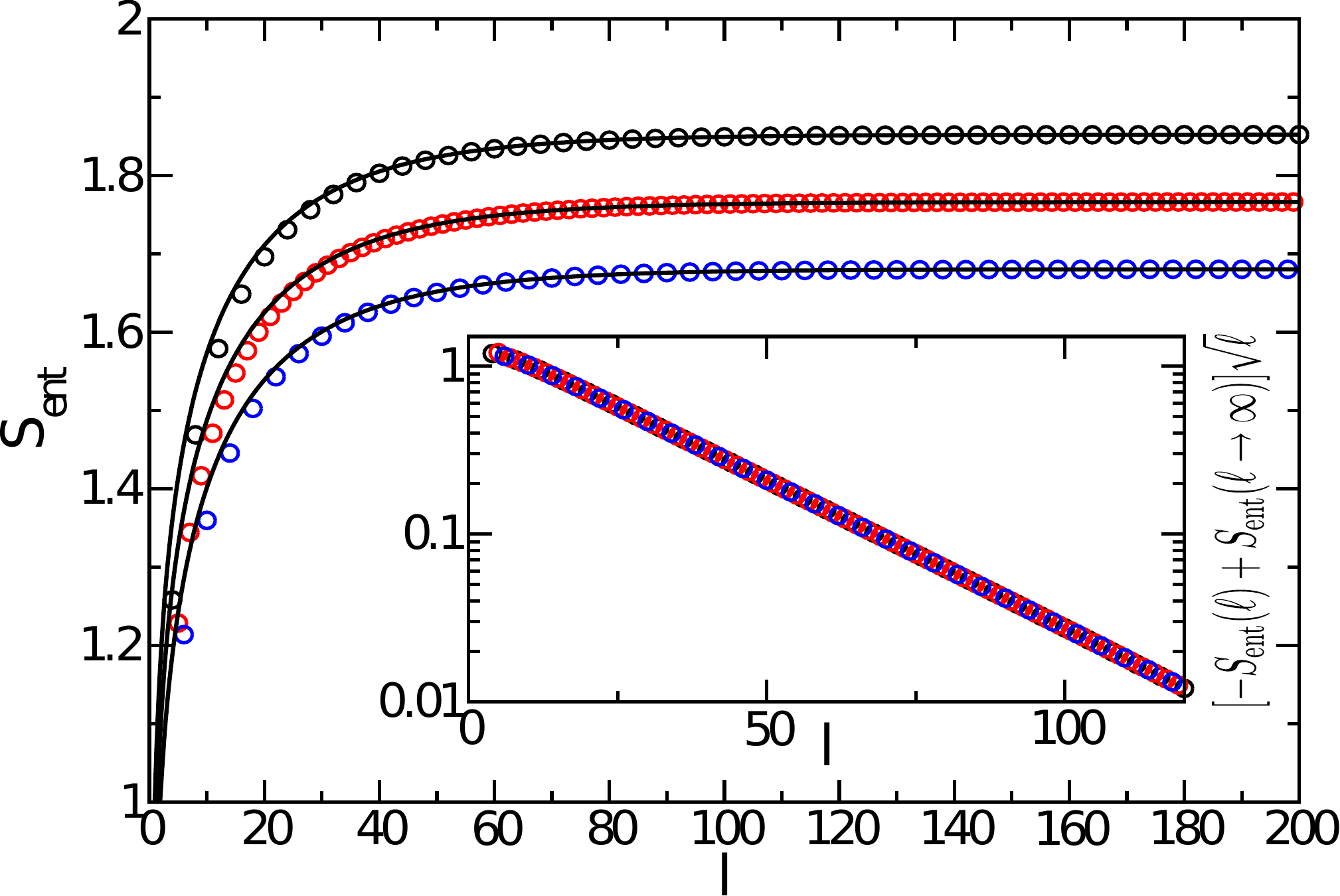}
\end{center}
\caption{$S_{\rm{ent}}$ (circles) for a block in an infinite chain at
  $|\delta|=0.02$ with (from bottom to top) $\ell$ even and
  $\delta<0$, $\ell$ odd, and $\ell$ even and $\delta>0$ compared to
  the scaling prediction (\ref{EE5_dim_model}) (lines). The inset
  shows that all three curves for $2\ell/\xi\gg 1$ collapse onto a
  single line given by the asymptotics of the Bessel function (see
  text).}
\label{Fig_Bessel}
\end{figure}

(3) Finally, there is also the regime $\xi\lesssim\ell$ where the
formula (\ref{EE5}) should hold which gives the leading correction to
the constant value obtained in the limit $\xi\ll\ell$. In the case of
the dimerized model considered here, the leading correlation length
$\xi$ will show up twice: once with wave vector $k=\pi/2$ and once
with wave vector $k=-\pi/2$ \cite{KluemperScheeren,SirkerKluemperEPL}.
Therefore, to leading order, we expect the entanglement entropy to
scale as
\begin{equation}
\label{EE5_dim_model}
S_{\rm{ent}}=S_{\rm{ent}}(\ell\to\infty)-\frac{1}{4}K_0(2\ell/\xi)
\end{equation}
with $\xi$ as given in equation~(\ref{Corrlength}). Fixing $S_{\rm
  ent}(\ell\to\infty)$ by the numerical data for large block sizes,
there is thus no free fitting parameter. The comparison in
figure~\ref{Fig_Bessel} with data for $|\delta|=0.02$, in which case
$\xi\approx 25$, confirms this scaling prediction.
For large arguments, the Bessel function asymptotically scales as
$K_0(x)\sim\sqrt{\frac{\pi}{2x}}\exp(-x)$. The inset of
figure~\ref{Fig_Bessel} demonstrates that this is indeed the correct
scaling of the entanglement entropy for large block sizes. The data
for the three different cuts, in particular, collapse onto a single
line after subtracting the constant part
$S_{\rm{ent}}(\ell\to\infty)$.

\subsection{A block at the end of a semi-infinite chain}
To calculate the entanglement entropy for a block which includes the
end of a semi-infinite chain, we can use the analytic results for the
single-particle eigenstates of a chain with OBC and $N$ odd given in
section~\ref{OBC_odd} and the numerical results for $N$ even from
section \ref{OBC_even}. In both cases, the matrix elements of the
correlation matrix $C_{ij}$, see equation~(\ref{Cij}), can be
expressed as
\begin{equation}
\label{Cij_2}
C_{ij}=\sum_{k\;\, \rm{occupied}} (\Psi^k_i)^* \Psi^k_j
\end{equation}
with the sum running over all occupied single-particle states.
Diagonalizing the correlation matrix then gives access to the
single-particle eigenenergies $\epsilon_k$ of the entanglement
Hamiltonian using the relation (\ref{EV_relation}).

\begin{figure}
\begin{center}
\includegraphics*[width=0.75\columnwidth]{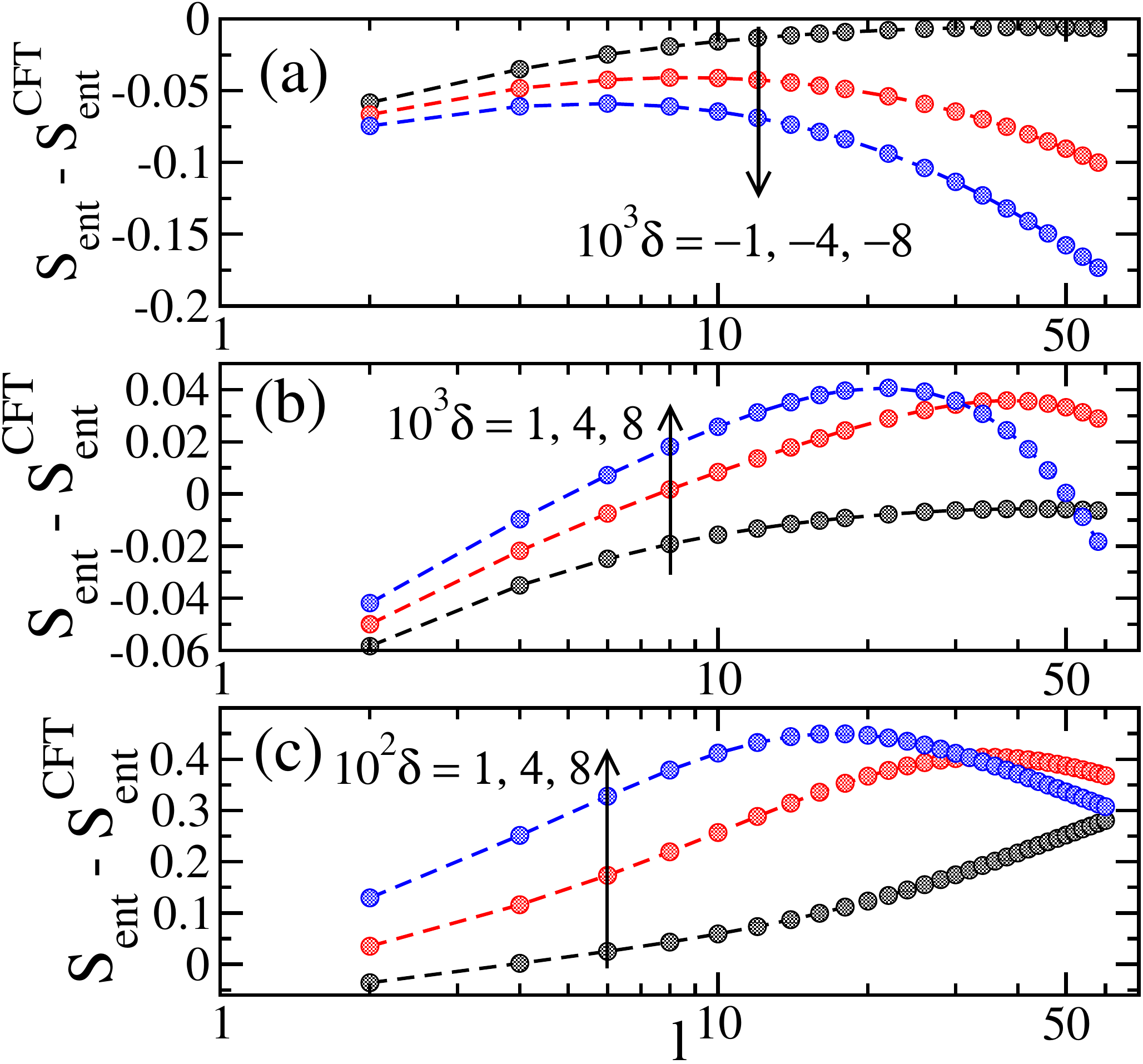}
\end{center}
\caption{$S_{\rm{ent}}$ for a block at the end of a semi-infinite
  chain at small dimerizations with $\ell$ even for (a) $\delta<0$
  where the boundary is trivial, (b) $\delta>0$ where a localized
  boundary state is present and filled, and (c) $\delta>0$ for $N$
  even where a localized boundary state is present but only
  half-filled. In all cases the CFT result (\ref{EE3}) in the
  non-dimerized case with $s_1=0.726$ has been subtracted. The dashed
  lines are a guide to the eye.}
\label{Fig_semi_inf0}
\end{figure}

Results for small dimerizations are shown in figure~\ref{Fig_semi_inf0}.
While the entanglement entropy for $\delta<0$, where a strong bond
terminates the semi-infinite chain, behaves qualitatively similar to
the case of a block in an infinite chain (see
figure~\ref{Sent_small_dim}) we find a much more complicated scaling
for $\delta>0$ and $N$ odd. In this case a localized boundary state is
present leading to a competition between an increase of the boundary
contribution to $S_{\rm{ent}}$ with increasing $\delta$ and a decrease
of the bulk contribution due to the decreasing correlation length
$\xi\sim 1/\delta$. As a result, the curves for different
dimerizations cross, see figure~\ref{Fig_semi_inf0}(b).

The final possibility is to consider the semi-infinite chain with a
localized edge state which is not fully occupied, as realized in a
chain $N$ even at half-filling, see figure \ref{Fig_semi_inf0}(c).  As
for the fully occupied edge state there is a competition between the
increasing boundary entropy and decreasing bulk entropy with $\delta$.
However, due to the higher degeneracy associated with the partially
occupied edge state, see also the entanglement spectra in the
following section, the curves for the same dimerizations will cross at
a much larger sub-block size, than for a fully occupied edge state.
Therefore we show in figure \ref{Fig_semi_inf0}(c) data for
dimerizations which are an order of magnitude larger than in the other
two cases shown in figure \ref{Fig_semi_inf0}(a,b).

\begin{figure}
\begin{center}
\includegraphics*[width=0.75\columnwidth]{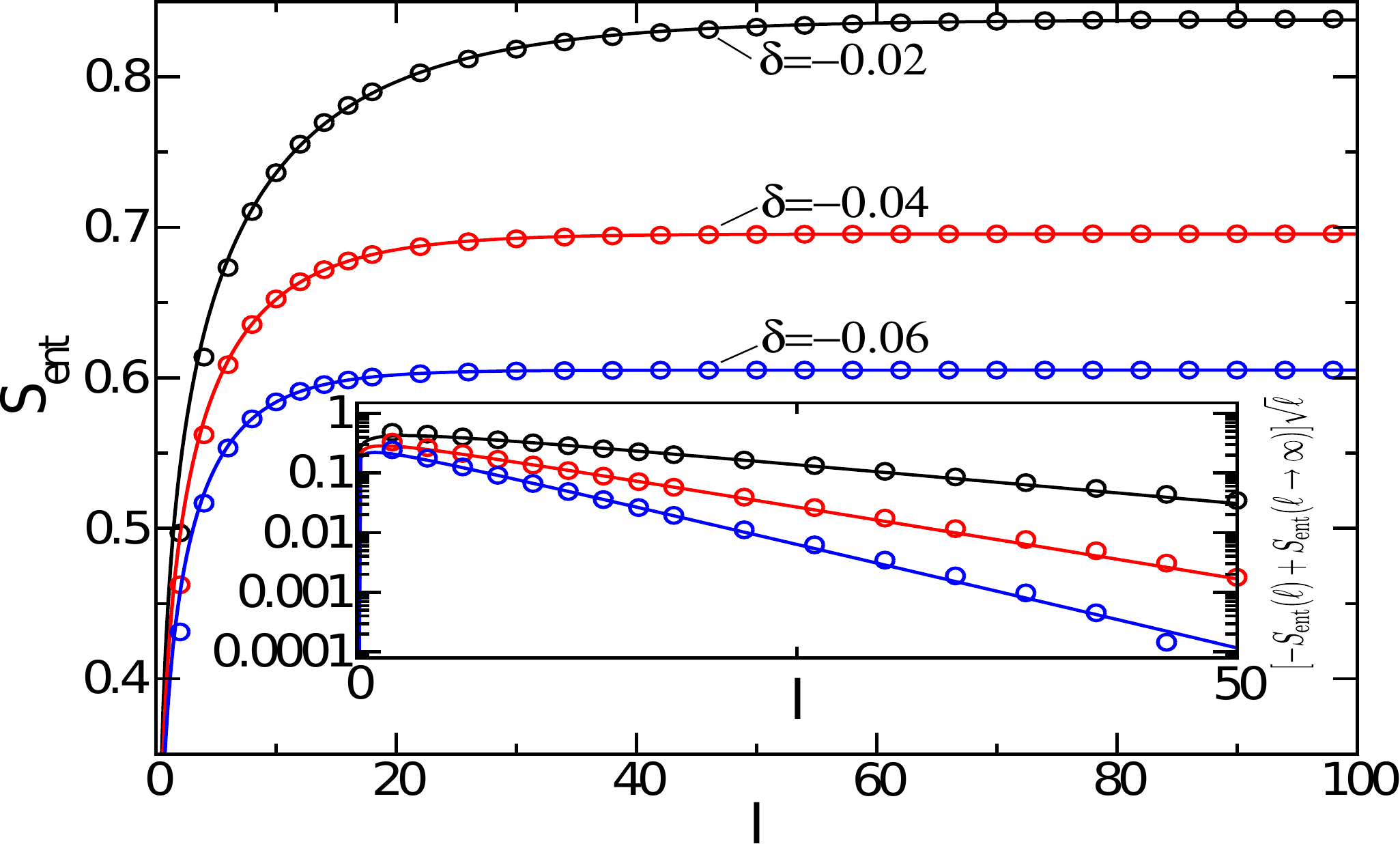}
\end{center}
\caption{$S_{\rm{ent}}$ for a block at the end of a semi-infinite
  chain with $\ell$ even, fitted by equation~(\ref{Bessel2}) with the
  fitting parameter $\tilde\xi$ given in table \ref{tab1}.  The inset
  shows that the scaling is still roughly consistent with the Bessel
  function asymptotics.}
\label{Fig_semi_inf1}
\end{figure}

For large block size $\ell$ and finite dimerization we again expect
that $S_{\rm{ent}}$ saturates. Let us first concentrate on the case
$\delta<0$ shown in figure~\ref{Fig_semi_inf1}.
In this case $S_{\rm{ent}}$ is monotonically increasing. We can thus
try to fit the data for large block sizes by
\begin{equation}
\label{Bessel2}
S_{\rm{ent}} = S_{\rm{ent}}(\ell\to\infty)-\frac{1}{8} K_0(2\ell/\tilde\xi)
\end{equation}
similar to the CFT result (\ref{EE5}) but with $\tilde\xi$ as a free
parameter. Doing so we obtain very good fits with parameters as given
in table \ref{tab1}.
\begin{table}
  \caption{\label{tab1} Parameters in equation~(\ref{Bessel2}) used to fit the data for large block sizes shown in figure~\ref{Fig_semi_inf1}.}
\begin{center}
\begin{indented}
\item[] \begin{tabular}{c|c|c|c|c}
$\delta$ & $S_{\rm{ent}}(\ell\to\infty)$ & $\tilde\xi$ & $\xi$ & $\xi/\tilde\xi$ \\ \hline
-0.02 & 0.838 & 34.80 & 49.99 & 1.44 \\ \hline
-0.04 & 0.696 & 17.59 & 24.99 & 1.42 \\ \hline
-0.06 & 0.605 & 11.89 & 16.65 & 1.40
\end{tabular}
\end{indented}
\end{center}
\end{table}
The correlation length $\tilde\xi$ obtained from the fits seems to be
about a factor $\sqrt{2}$ smaller than the bulk correlation length
$\xi$ given in equation~(\ref{Corrlength}). Note, however, that the
heuristic fitting formula (\ref{Bessel2}) is too naive to be able to
fully capture the behaviour of $S_{\rm ent}$. A two-point correlation
function near a boundary does not only depend on the distance between
the two operators but also on the distance of the operators from the
boundary so that a third length scale has to appear.

\begin{figure}
\begin{center}
\includegraphics*[width=0.75\columnwidth]{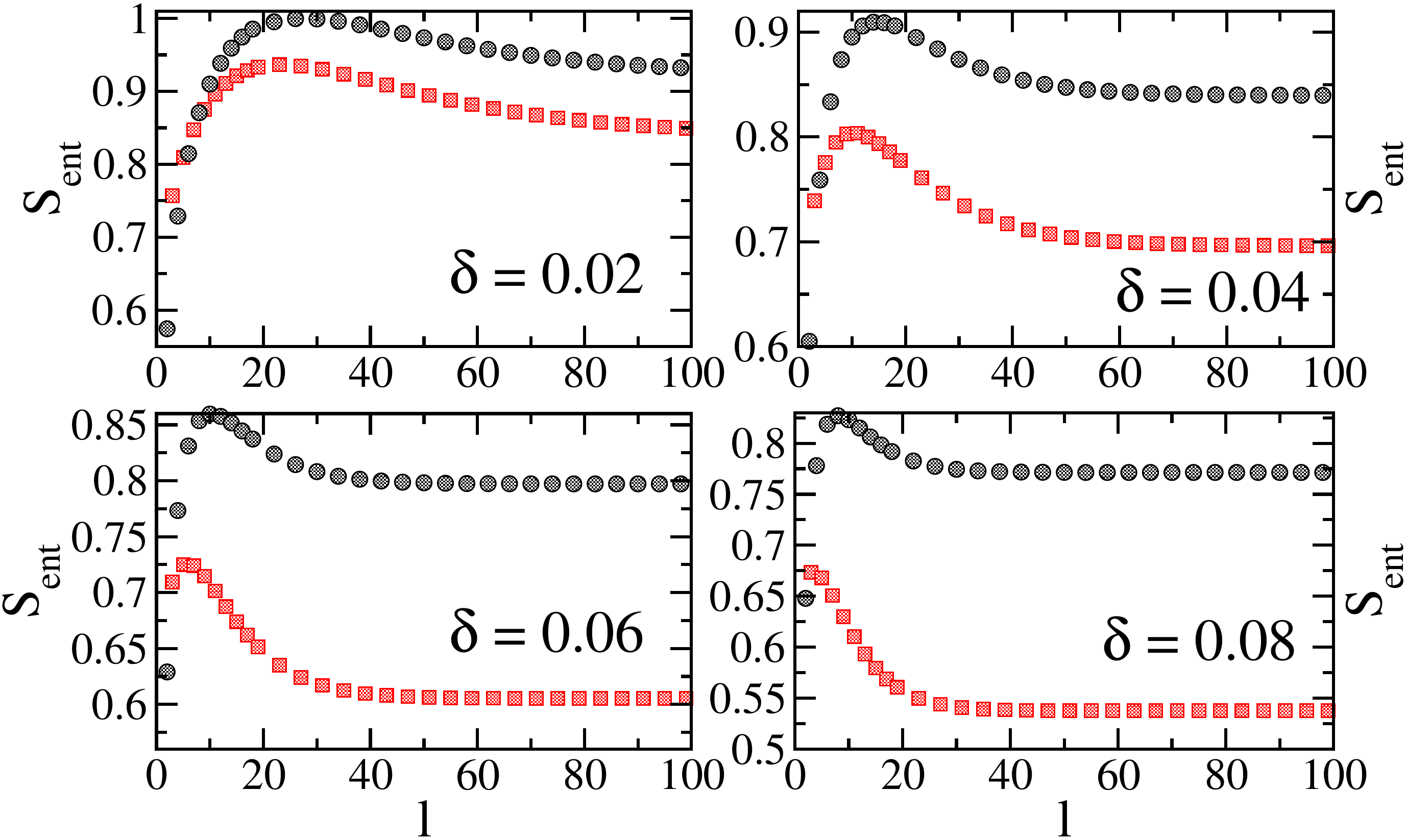}
\end{center}
\caption{$S_{\rm{ent}}$ for a block at the end of a semi-infinite
  chain with $\delta>0$ and a fully occupied boundary state. Shown are
  data for even (circles) and odd block sizes (squares).}
\label{Fig_semi_inf2}
\end{figure}

Let us now return to the case $\delta>0$ with the fully occupied
boundary state where figure~\ref{Fig_semi_inf0}(b) already indicates
that the scaling behaviour of the entanglement entropy with block size
is completely changed. Figure~\ref{Fig_semi_inf2} shows that the
competition between boundary and bulk contributions leads to a maximum
in $S_{\rm{ent}}$ which becomes sharper and whose position shifts to
smaller block sizes with increasing dimerization.
Consequently, the constant value $S_{\rm{ent}}(\ell\to\infty)$ is
now approached from above. Clearly a scaling as in (\ref{Bessel2}) is
no longer valid. 

\begin{figure}
\begin{center}
\includegraphics*[width=0.75\columnwidth]{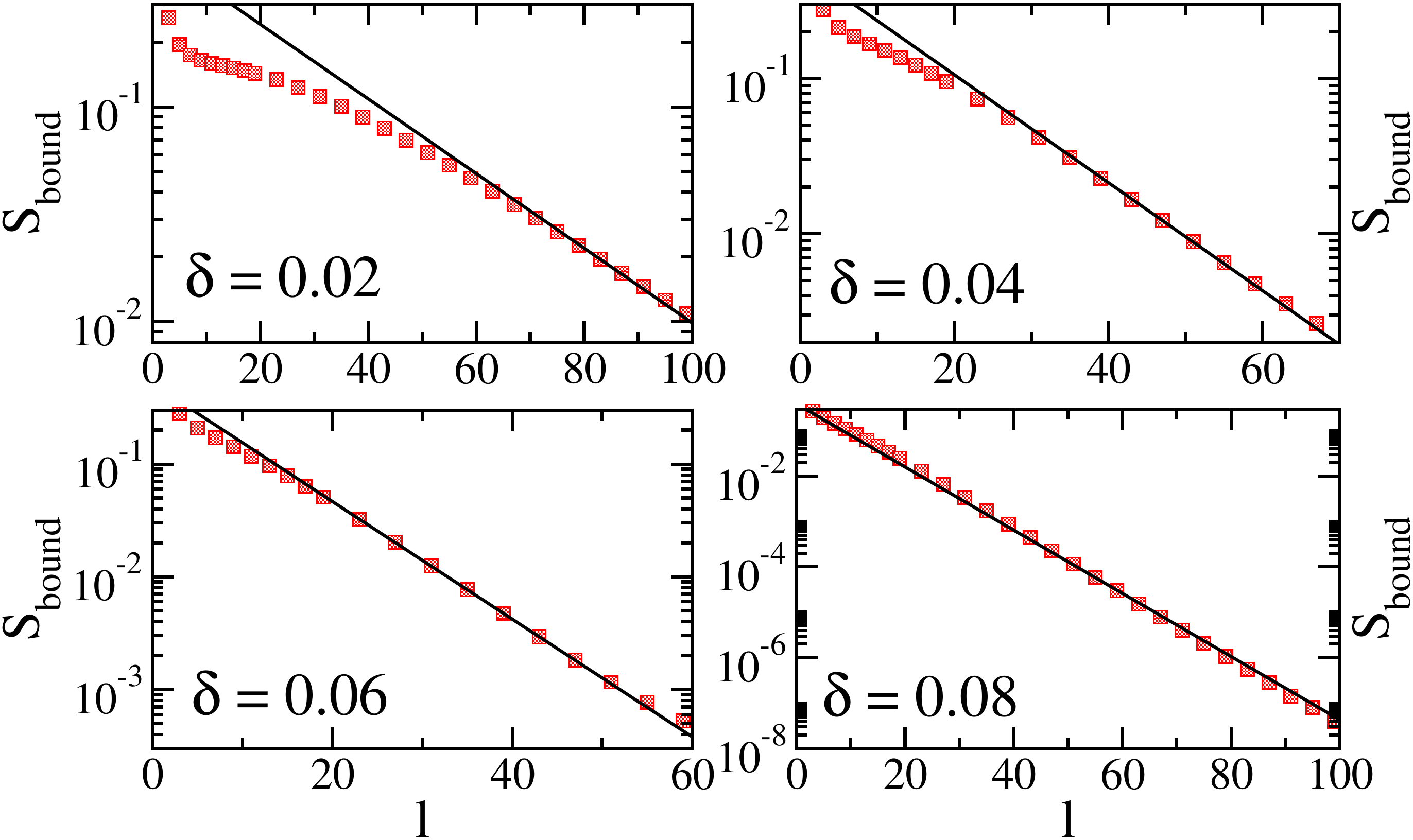}
\end{center}
\caption{$S_{\rm{bound}}$ as defined in equation (\ref{Sbound}) for
  different $\delta>0$ where the boundary state is fully occupied. The
  lines display $S_{\rm bound}=a\exp(-\ell/\xi_{\rm loc})$ with the
  localization length (\ref{loc_length}) and $a$ used as the only
  fitting parameter.}
\label{Fig_semi_inf3}
\end{figure}

To verify that the additional entanglement entropy is indeed caused by
the boundary state we can define
\begin{equation}
\label{Sbound}
S_{\rm bound}(\ell) = S^{\delta>0}_{\rm ent}(\ell,\, \ell\, \rm{odd}) - S^{\delta<0}_{\rm ent}(\ell+1,\, \ell+1 \, \rm{even})\; .
\end{equation}
For both entropies in equation (\ref{Sbound}) a weak bond is cut so
that for large block sizes $\ell$ any remaining difference is directly
related to the localized boundary state present for $\delta>0$. Figure
\ref{Fig_semi_inf3} shows that $S_{\rm bound}(\ell)$ is exponentially
decaying with block size.
According to equation~(\ref{N_odd_zero}) a non-trivial boundary leads
to an exponentially localized zero energy state with a localization
length
\begin{equation}
\label{loc_length}
\xi^{-1}_{\rm{loc}}=\ln\left(\frac{1+\delta}{1-\delta}\right) .
\end{equation}
The exponential fits in figure \ref{Fig_semi_inf3} using this
correlation length are indeed in excellent agreement with the data for
$S_{\rm bound}(\ell)$ confirming that the boundary state leads to
additional entanglement which is exponentially localized at the
boundary.

\begin{figure}
\begin{center}
\includegraphics*[width=0.75\columnwidth]{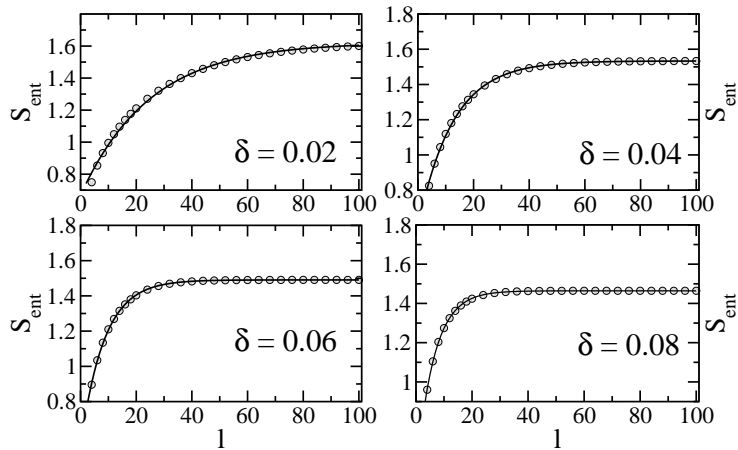}
\end{center}
\caption{$S_{\rm{ent}}$ for $\delta>0$ and a partially filled boundary
  state. Shown are data for even block sizes (circles). The lines are
  exponential fits, see equation (\ref{Ent_exp}).}
\label{Fig_semi_inf4}
\end{figure}

\begin{figure}
\begin{center}
\includegraphics*[width=0.75\columnwidth]{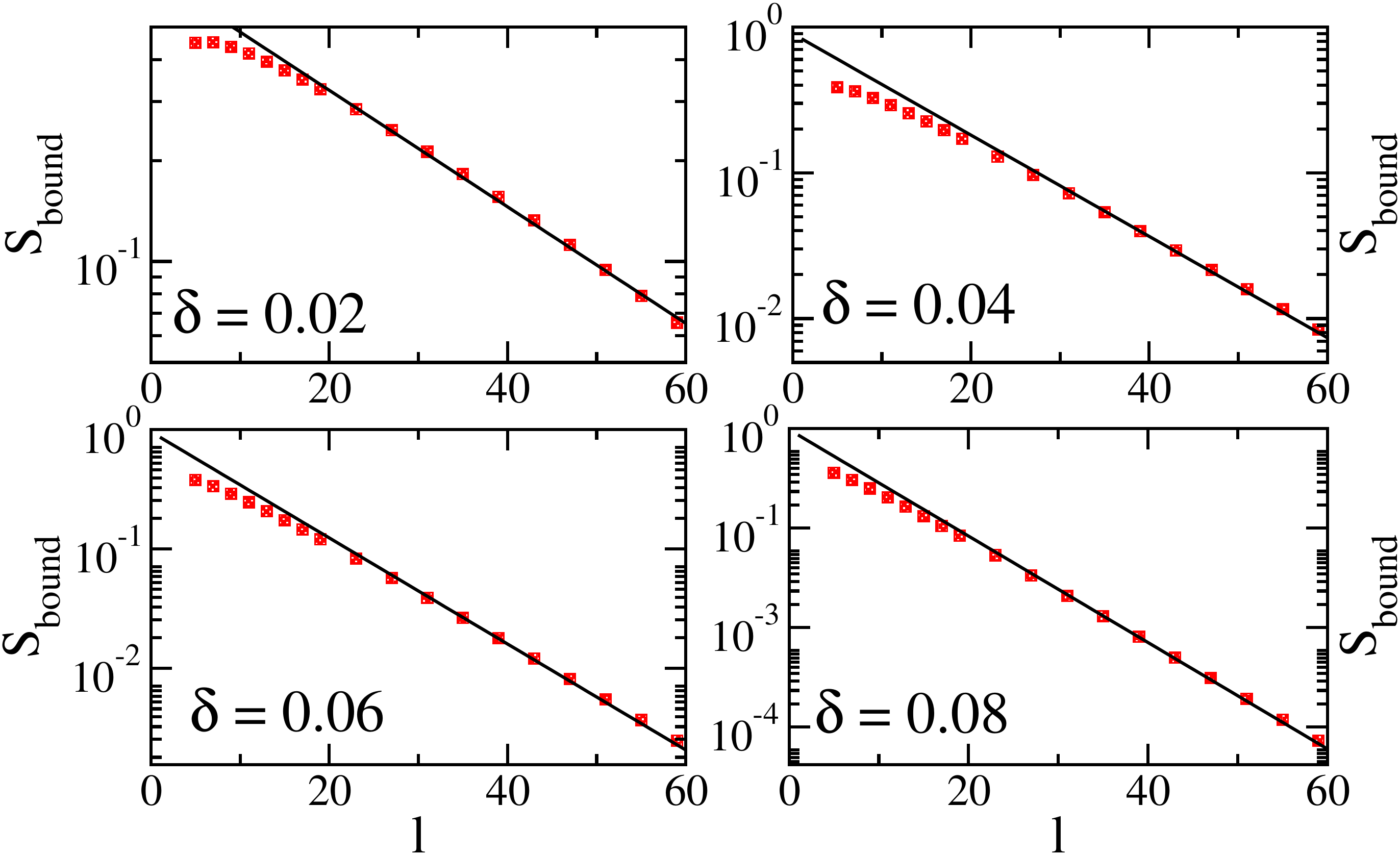}
\end{center}
\caption{$S_{\rm{bound}}$ as defined in equation (\ref{Sbound}) for
  $\delta>0$ with a partially filled boundary state. The lines display
  $S_{\rm bound}=a\exp(-\ell/\xi_{\rm loc})$ with the localization
  length (\ref{loc_length}) and $a$ used as the only fitting
  parameter.}
\label{Fig_semi_inf5}
\end{figure}

For $\delta>0$, $N$ even, and half-filling the boundary state is only
half-filled. As shown in figure \ref{Fig_semi_inf4} this qualitatively
changes the behaviour of $S_{\rm{ent}}$ as compared to the case where
the edge state is completely filled. In particular, there is no longer
a maximum yet the scaling with block size is also not described by a
modified Bessel function as was the case for $\delta<0$.
Instead, the scaling seems purely exponential
\begin{equation}
\label{Ent_exp}
S_{\rm{ent}}(\ell)=S_{\rm{ent}}(\ell \rightarrow\infty)-a\exp(-2\ell/\xi)
\end{equation}
with $\xi$ being the leading correlation length, see equation
(\ref{Corrlength}), and $a$ a fitting parameter. The boundary
contribution of the partially filled edge state can again be
calculated from equation (\ref{Sbound}) and is exponentially decaying
with block size similar to the case of the completely filled edge
state, see figure \ref{Fig_semi_inf5}.
To summarize, we have shown that the presence of localized boundary
states in a massive phase completely changes the scaling of
$S_{\rm{ent}}(\ell)$ with block size $\ell$ for a block at the end of
a semi-infinite chain. The boundary state leads, in particular, to
additional entanglement which is exponentially loacalized. Further
insight into the different entanglement properties of a block with or
without a localized boundary state can be gained by directly
investigating the entanglement spectrum.

\section{Entanglement spectra}
\label{Entanglement_spectra}
Instead of just considering the entanglement entropy, which is a map
from the $2^\ell$ dimensional space of eigenvalues of $\rho_A$ into
the real numbers, it is often instructive to consider the spectrum of
$\rho_A$---the so-called entanglement spectrum---directly.

\subsection{A block in an infinite chain}
\begin{figure}
\begin{center}
\includegraphics*[width=0.99\columnwidth]{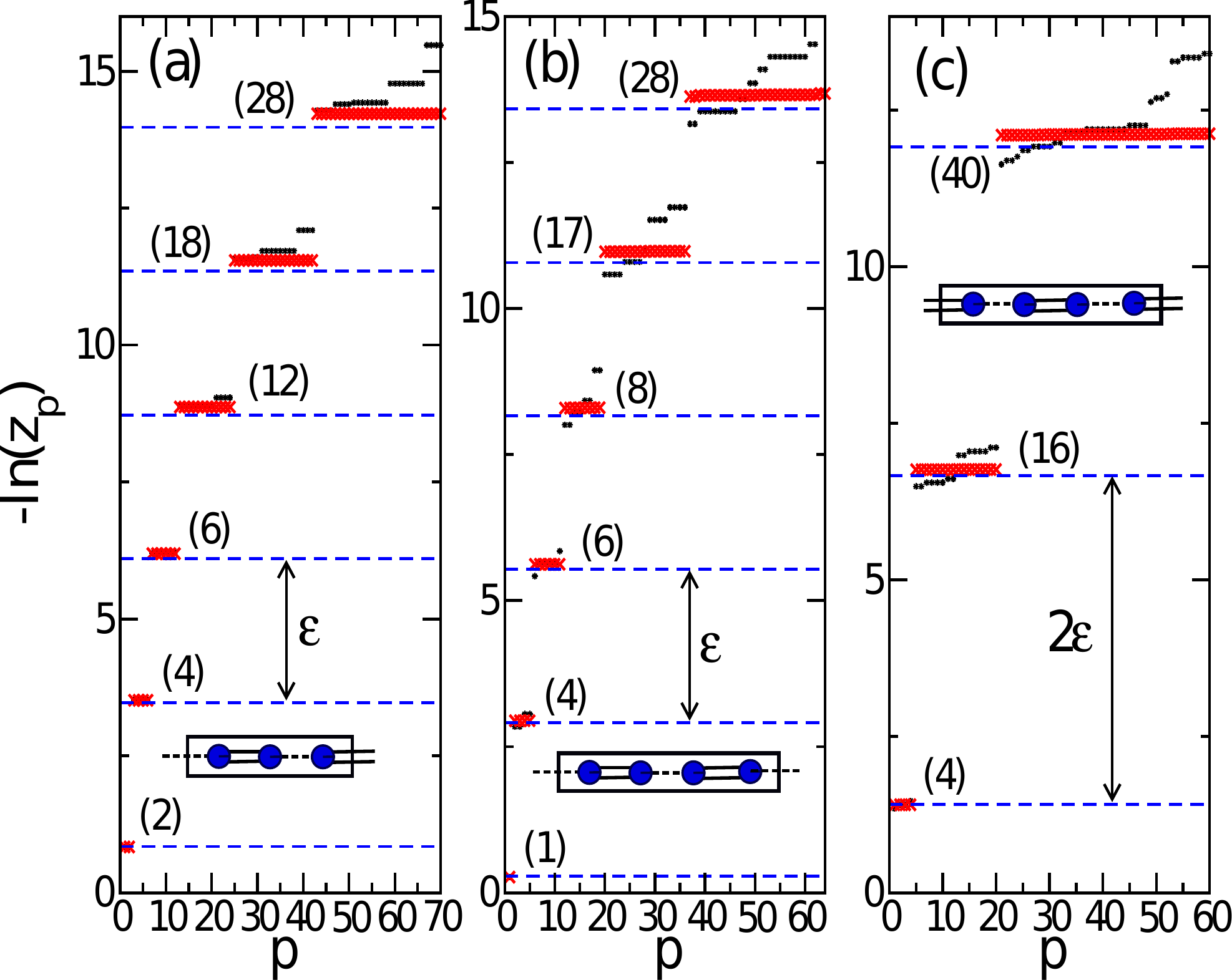}
\end{center}
\caption{Entanglement spectrum for a block in an infinite chain at
  $|\delta|=0.1$ with (a) odd block size, (b) even block size cutting
  two weak bonds, and (c) even block size cutting two strong bonds.
  The black stars are data for $\ell =19$ ($\ell=20$), the red crosses
  for $\ell=69$ ($\ell=70$).  The dashed lines are the single particle
  levels in the thermodynamic limit, the numbers in bracket give the
  degeneracy $d_k$ of each level.}
\label{Fig_EntSpectrum_inf}
\end{figure}

As for the entanglement entropy we can use the analytical result for
the elements of the correlation matrix in the thermodynamic limit to
determine the spectrum of the reduced density matrix. In
figure~\ref{Fig_EntSpectrum_inf} the logarithm of the eigenvalues $z_p$
of the reduced density matrix, see equation~(\ref{EV_rho}), for the three
different cuts possible are shown.
The pictograms in the figure show whether the block size is even or
odd (but not the actual block size!) and if strong (double lines) or
weak (single dashed lines) are cut. While the structure is complicated
for finite block sizes $\ell$, the distribution of eigenvalues becomes
very simple in the limit $\ell\to\infty$. In this case, the single
particle energies $\epsilon_k$ of the entanglement Hamiltonian
(\ref{rho_red}) become equally spaced and the level spacing $\epsilon$
can be obtained analytically using corner transfer methods
\cite{PeschelEisler,SirkerEntanglement}. One finds
\begin{equation}
\label{level_spacing}
\epsilon=\pi I(x')/I(x) \quad \mbox{with} \quad x=(1-\delta)/(1+\delta),
\end{equation}
$x'=\sqrt{1-x^2}$, and $I(x)$ being the elliptic integral of the first
kind. This is the correct level spacing for odd block size as well as
for even block size if two weak bonds are cut. Cutting two strong
bonds, on the other hand, leads to a level spacing $2\epsilon$, see
figure~\ref{Fig_EntSpectrum_inf}.

The eigenvalues $z_p$ of the reduced density matrix in the
thermodynamic limit can then be obtained by filling up the single
particle levels. The dominant eigenvalue of $\rho_A$ is obtained by
filling up the `Fermi sea' of negative eigenvalues $\epsilon_k$. In
the case shown in figure~\ref{Fig_EntSpectrum_inf}(b)---where the
block size is even and two weak bonds are cut---this leads to a unique
dominant eigenvalue. In the case where either one
[figure~\ref{Fig_EntSpectrum_inf}(a)] or two
[figure~\ref{Fig_EntSpectrum_inf}(c)] strong bonds are cut the lowest
eigenvalue is, however, degenerate. This can be understood as follows:
Cutting a strong bond leads to an edge spin. For the case $\ell$ odd
where one strong bond is cut this leads to an eigenenergy
$\epsilon_{(\ell+1)/2}\equiv 0$.  This level can now either be empty
or filled leading to a two-fold degeneracy of the leading eigenvalue.
For the case $\ell$ even where two strong bonds are cut, eigenvalues
$\epsilon_{\ell/2}=-\epsilon_{\ell/2+1}$ appear which approach zero in
the thermodynamic limit. Consequently, the dominant eigenvalue of
$\rho_A$ becomes four-fold degenerate. Thus the low-energy
entanglement spectrum allows to directly identify the number of `edge
states' caused by cutting a strong bond.

Equation~(\ref{level_spacing}) only gives the level spacing. In order
to obtain the allowed values for the eigenvalues $z_p$ of $\rho_A$ one
has to use, in addition, the normalization condition
\begin{equation}
\label{Normalization}
\!\!\!\!\!\!\!\!\!\!\!\!\!\!\! 1=\Tr\rho_A=\sum_p z_p =\sum_{k=0}^\infty d_k \e^{-\epsilon_0-k\epsilon}\;\; \Rightarrow \;\;\epsilon_0=\ln\left(\sum_{k=0}^\infty d_k \e^{-k\epsilon}\right) \; .
\end{equation}
Here $d_k$ is the number of degenerate eigenvalues $\{z_p\}$ for which
$z_p=\exp(-\epsilon_0-k\epsilon)$ with $k$ fixed. The levels
$\epsilon_0+k\epsilon$ or $\epsilon_0+2k\epsilon$, respectively, with
$k=0,1,2,3,\cdots$ are shown as dashed lines in
figure~\ref{Fig_EntSpectrum_inf}.

\subsection{A block at the end of a semi-infinite chain}
While the case of a block in an infinite chain is well understood, the
interesting case of the entanglement spectrum for a block at the end
of a dimerized semi-infinite chain with or without a localized
boundary state has not been considered so far.  We have already shown
that a localized boundary state causes additional entanglement which
is also localized at the boundary, see figure \ref{Fig_semi_inf3}.
Here we want to further investigate this case by directly looking at
the entanglement spectrum.

\begin{figure}
\begin{center}
\includegraphics*[width=0.99\columnwidth]{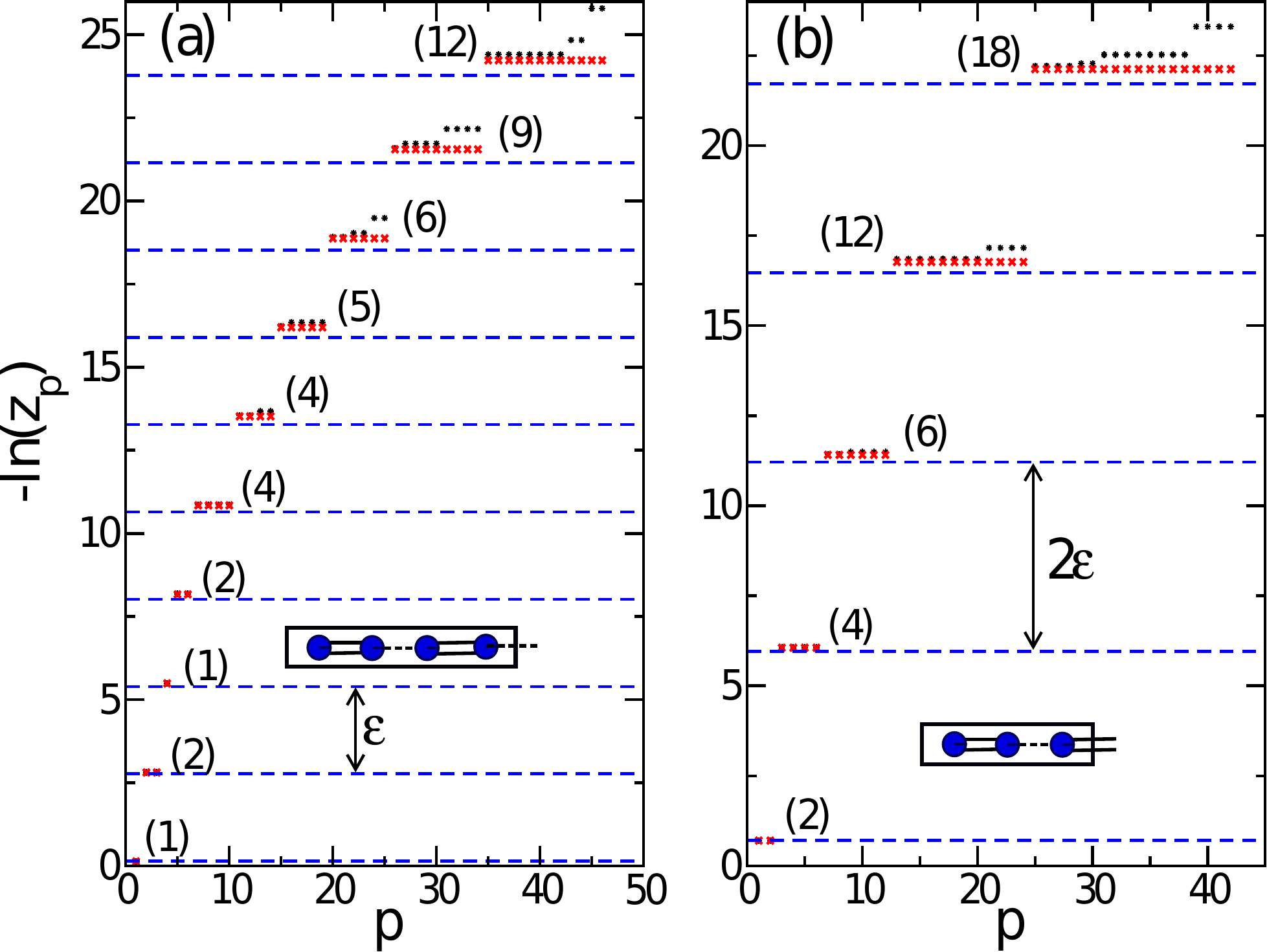}
\end{center}
\caption{Entanglement spectrum for a block at the end of a
  semi-infinite chain at $\delta=-0.1$ with (a) even block size, and
  (b) odd block size.  The black stars are data for $\ell =19$
  ($\ell=20$), the red crosses for $\ell=69$ ($\ell=70$).  The dashed
  lines are the single particle levels in the thermodynamic limit, the
  numbers in bracket give the degeneracy of each level.}
\label{Fig_EntSpectrum_semi_inf1}
\end{figure}

For $N$ odd, we can distinguish four cases now: $\delta<0$
(topological trivial case) or $\delta>0$ (SPT case) with $\ell$ either
even or odd. In addition, we have to consider the two cases $\ell$
even or odd with $N$ even and $\delta>0$, where the edge state is only
partially filled, separately. We start with the topological trivial
case $\delta<0$. Data for even and odd block sizes with $\delta=-0.1$
are shown in figure~\ref{Fig_EntSpectrum_semi_inf1}.
For even block sizes, see figure~\ref{Fig_EntSpectrum_semi_inf1}(a),
the leading eigenvalue is unique and the separation of the
single-particle levels is given by equation~(\ref{level_spacing}) as
in the case of an even block in an infinite chain where two weak bonds
are cut, see figure~\ref{Fig_EntSpectrum_inf}(b). However, the
degeneracy structure of the higher levels is different which explains
the differences in the entanglement entropy discussed in the previous
subsection. For the case of an odd block size, shown in
figure~\ref{Fig_EntSpectrum_semi_inf1}(b), the degeneracy structure of
the spectrum is the same as for a block of odd length in an infinite
chain, see figure~\ref{Fig_EntSpectrum_inf}(a), however, the level
spacing in the semi-infinite chain is $2\epsilon$ instead of
$\epsilon$ for the infinite chain. Note that the finite size
corrections for a block at the end of a semi-infinite chain in the
topological trivial case are of similar magnitude as for the infinite
chain, i.e., for the lowest single-particle levels the data for the
smaller block sizes $\ell=19,20$ and the larger ones, $\ell=69,70$,
hardly deviate from each other.

\begin{figure}
\begin{center}
\includegraphics*[width=0.99\columnwidth]{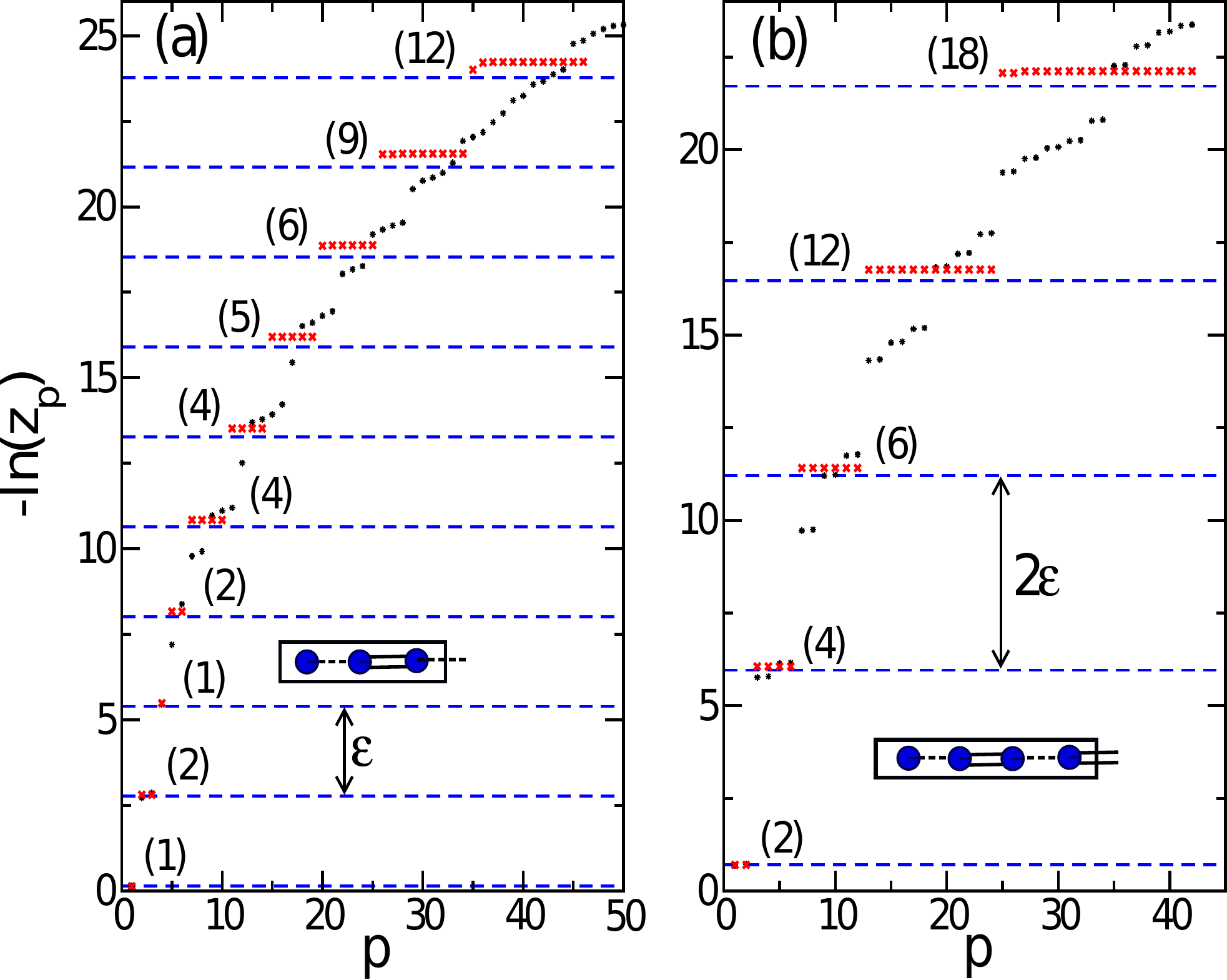}
\end{center}
\caption{Entanglement spectrum for a block at the end of a
  semi-infinite chain at $\delta=0.1$ with a fully occupied boundary
  state for (a) odd block size, and (b) even block size.  The black
  stars are data for $\ell =19$ ($\ell=20$), the red crosses for
  $\ell=69$ ($\ell=70$).  The dashed lines are the single particle
  levels in the limit $\ell\to\infty$, the numbers in bracket give the
  degeneracy of each level.}
\label{Fig_EntSpectrum_semi_inf2}
\end{figure}

This is completely different in the SPT phase---the case $\delta=0.1$
with $N$ odd is shown as example in
figure~\ref{Fig_EntSpectrum_semi_inf2}---where finite size corrections
are much larger.
In the limit $\ell\to\infty$, on the other hand, both the level
spacing and the degeneracies agree between the cases $\delta<0$,
$\ell$ even [figure~\ref{Fig_EntSpectrum_semi_inf1}(a)] and
$\delta>0$, $\ell$ odd [figure~\ref{Fig_EntSpectrum_semi_inf2}(a)] as
well as $\delta<0$, $\ell$ odd
[figure~\ref{Fig_EntSpectrum_semi_inf1}(b)] and $\delta>0$, $\ell$
even [figure~\ref{Fig_EntSpectrum_semi_inf2}(b)]. This means that the
localized boundary state present for $\delta>0$ does not play any role
in the limit $\ell\to\infty$ and that the spectra in the trivial and
the SPT phase become identical in this limit.

The dramatic change in the entanglement spectra for small block sizes
if a localized state is present---which leads to the non-monotonic
behaviour of the entanglement entropy shown in
figure~\ref{Fig_semi_inf2}---can be understood as follows: The
non-trivial edge leads to an exponentially localized zero energy
state, equation~(\ref{N_odd_zero}), with a localization length as
given in equation (\ref{loc_length}).

For the case $\delta=0.1$ considered here the localization length is
$\xi_{\rm{loc}}\approx 1/2\delta= 5$ lattice sites. The low-energy
part of the entanglement spectrum will thus become similar to the one
without the localized state if $\ell\gg\xi_{\rm{loc}}$ while it will
be substantially modified for $\ell\lesssim \xi_{\rm{loc}}$. The
modification of the entanglement spectrum and the entanglement entropy
is thus an effect of finite block size which is directly tied to the
appearance of localized states in the SPT phase.

\begin{figure}
\begin{center}
\includegraphics*[width=0.99\columnwidth]{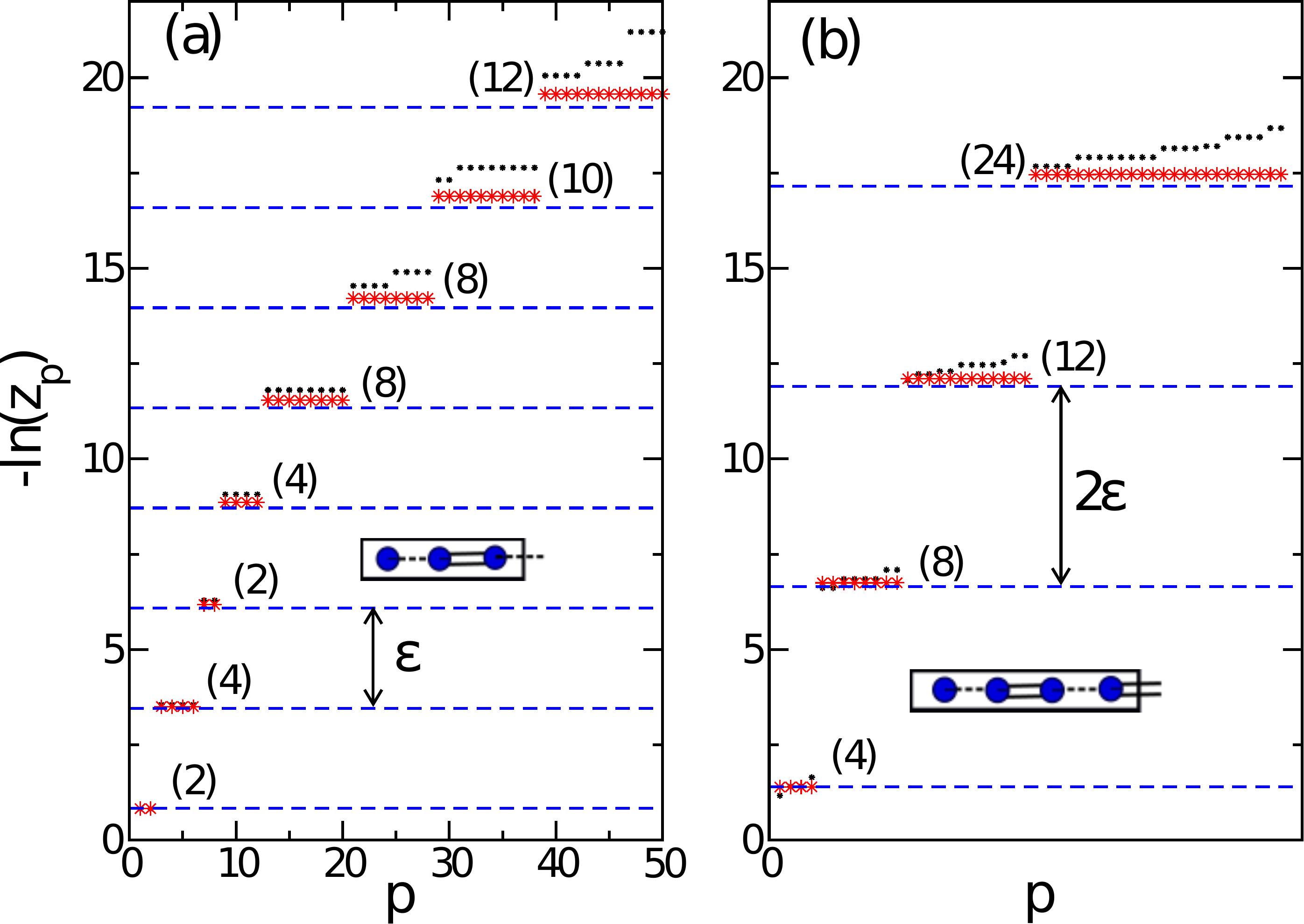}
\end{center}
\caption{Entanglement spectrum for a block at the end of a
  semi-infinite chain with $N$ even at $\delta=0.1$ (partially filled
  boundary state) with (a) odd block size, and (b) even block size.
  The black stars are data for $\ell =19$ ($\ell=20$), the red crosses
  for $\ell=69$ ($\ell=70$).  The dashed lines are the single particle
  levels in the thermodynamic limit, shifted by $\ln 2$ relative to
  the odd system size case due to the doubling in degeneracy of each
  level, see the numbers in parenthesis.}
\label{Fig_EntSpectrum_semi_inf3}
\end{figure}

The final possibility is to consider the case $N$ even, $\delta>0$
where the boundary state is only partially filled, see figure
\ref{Fig_EntSpectrum_semi_inf3}. In this case the degeneracy of the
levels is doubled compared to the case with a fully occupied boundary
state, shown in \ref{Fig_EntSpectrum_semi_inf2}. Furthermore,
corrections to the $\ell\to\infty$ limit for finite blocks are
smaller, explaining the different behaviour seen in the entanglement
entropy.

To summarize, we have shown that the presence of a localized boundary
state significantly alters the entanglement spectra for small block
sizes. In the limit $\ell\to\infty$, on the other hand, a free fermion
spectrum with an equidistant level spacing is recovered. While
creating virtual edges by considering a block in an infinite chain
gives the same degeneracy of the lowest level than a physical edge for
a corresponding block at the end of a semi-infinite chain, the
degeneracies of higher levels as well as the level spacing are, in
general, different. This explains the different scaling properties of
$S_{\rm{ent}}(\ell)$ for a block at the end as compared to a block in
the middle.

\section{The SSH model with nearest-neighbour repulsion}
\label{Interacting}
Finally, we will investigate the effects of interactions by adding a
nearest-neighbour repulsion term, see the Hamiltonian in equation
(\ref{SSH}). It has been shown that topological phases of interacting
one-dimensional quantum systems can be classified by their
entanglement properties
\cite{FidkowskiKitaev,PollmannTurner,TurnerPollmann}. Interaction
effects in one-dimensional dimerized fermionic systems with SPT phases
have already been studied in Ref.~\cite{ManmanaEssin}. However, in
this work a spinful dimerized Hubbard model was considered where for
$\delta\to 1$ (SPT phase), even in the limit of infinite onsite
Hubbard repulsion, one can still remove an electron from the edge site
without any energy cost. Thus the cases $\delta<0$ and $\delta>0$
remain topologically distinct in the presence of a Hubbard
interaction. For the spinless model with nearest-neighbour repulsion
considered here, the situation is different. For $U\to\infty$ the
system will be forced into a charge-density-wave (CDW) state with
every second site occupied so that, even for $\delta\to 1$, the edges
are no longer decoupled from the rest of the chain. We therefore
expect a phase transition in the model (\ref{SSH}) as a function of
interaction $U$ where the fidelity and the entanglement measures will
change. We focus first on open chains with an even number of sites and
use exact diagonalization and Arnoldi algorithms to calculate the
ground state properties for system sizes up to $N= 24$. In order to
firmly establish the change of the entanglement properties at the
phase transition we, furthermore, present light cone renormalization
group (LCRG) \cite{EnssSirker} results for infinite chains.

\subsection{Fidelity susceptibility}
\begin{figure}
\begin{center}
\includegraphics*[width=0.85\columnwidth]{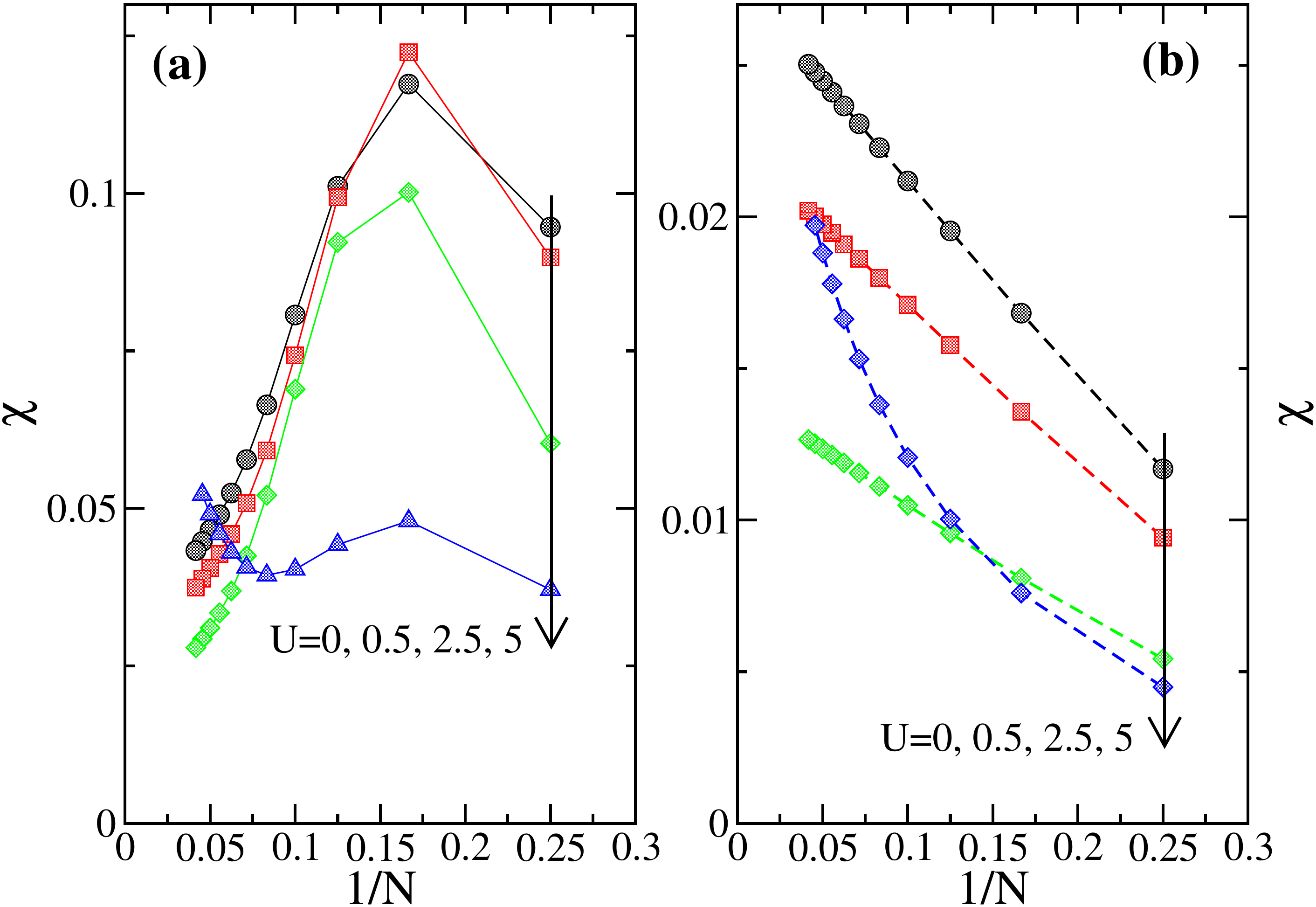}
\end{center}
\caption{Fidelity susceptibility for the interacting SSH chain
  (\ref{SSH}) with an even number of sites where (a) $\delta=0.5$, and
  (b) $\delta=-0.5$.}
\label{Fig_Fidelity_scaling_d_05_diff_U}
\end{figure}

In figure~\ref{Fig_Fidelity_scaling_d_05_diff_U} we show the fidelity
susceptibility at different interaction strengths $U$ for $\delta=0.5$
[topological phase for $U=0$] and $\delta=-0.5$ [trivial phase for
$U=0$].
For $U=0.5$ and $U=2.5$ the fidelity susceptibility behaves
qualitatively similar to the non-interacting case with the bulk
susceptibility $\chi_0$ being reduced with increasing interaction
strength. Furthermore, we still find $\chi_B(\delta<0)<0$ and
$\chi_B(\delta>0)>0$ with $\chi_B$ given by the slope of the curves in
figure~\ref{Fig_Fidelity_scaling_d_05_diff_U} for $1/N$ small. This is
an indication that the two cases remain topologically distinct even
for intermediate interaction strengths. For $U=5$, however, a drastic
change is observed. In this case $\chi_B$ is negative {\it both} for
$\delta<0$ and for $\delta>0$. Apparently the boundaries now behave
qualitatively similar suggesting that a phase transition has occurred
and edge states no longer exist. We will provide further evidence that
this is indeed the case in the following. $\chi_B$ can thus be used as
a measure to distinguish between phases with and without edge states
also in interacting systems.

\subsection{Entanglement spectra}
\begin{figure}
\begin{center}
\includegraphics*[width=0.99\columnwidth]{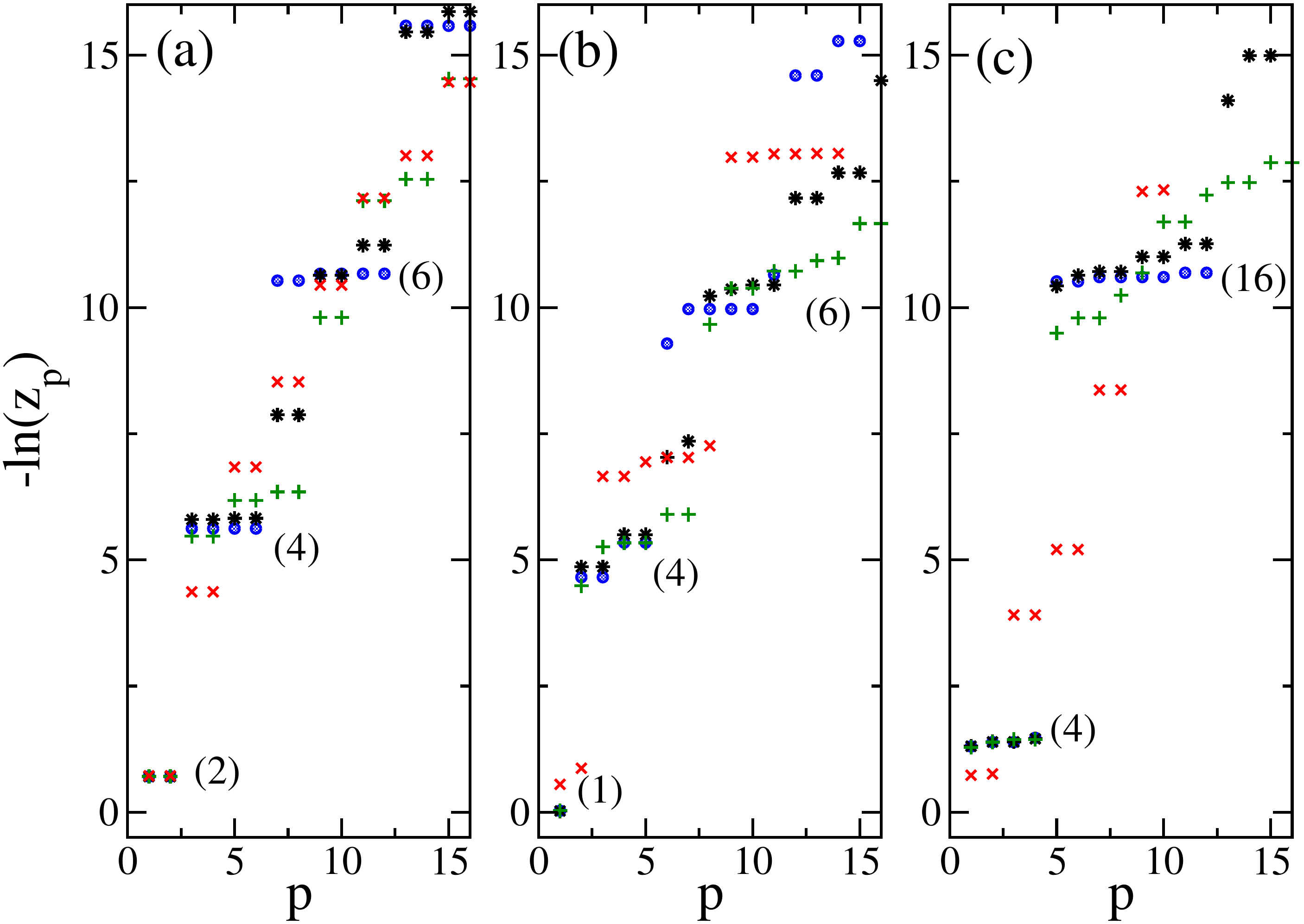}
\end{center}
\caption{Entanglement spectrum for a block centered at the middle of a
  chain with length $N=20$ at $|\delta|=0.5$ with (a) odd block size
  $\ell=5$, (b) even block $\ell=4$ size cutting two weak bonds, and
  (c) cutting two strong bonds.  The blue circles are data for $U=0$,
  the black stars for $U =0.5$, the green pluses for $U=2$, and the
  red crosses for $U=10$.  The numbers in parenthesis give the
  degeneracy of the levels for the non-interacting system in the limit
  $\ell\to\infty$.} 
\label{Fig_Int_EntSpectrum_Middle}
\end{figure}

We now turn to the entanglement spectra in the interacting case. We
concentrate on large dimerizations (as an example, data for
$|\delta|=0.5$ are shown throughout this section) where the
correlation length is small so that effects caused by the finite chain
length can be neglected for the low-energy part of the spectrum. As in
the non-interacting case, we start by choosing a block from the middle
of the chain, cutting (a) a weak and a strong bond, (b) two weak
bonds, and (c) two strong bonds, see figure
\ref{Fig_Int_EntSpectrum_Middle}.
For weak and intermediate interaction strengths, $U=0.5$ and $U=2$
respectively, the degeneracy of the lowest level is not altered while
for $U=10$ a twofold degeneracy starts to emerge in all three cases.
This can be understood as follows: For $U/t\to\infty$ the ground state
of the chain will become a superposition of the two degenerate product
states, $|\Psi\rangle = (|1010\cdots 10\rangle + |0101\cdots
01\rangle)/\sqrt{2}$, i.e., the system is driven into a symmetrized
CDW state ('cat state'). The reduced density matrix for any cut has
then two eigenvalues $1/2$ while all other eigenvalues are zero. The
corrections for non-zero $t/U\ll 1$ can be calculated perturbatively
similar to the case of the XXZ chain considered in
Refs.~\cite{AlbaHaque,SirkerEntanglement}. The entanglement spectrum
will then become equidistant again---though not all levels are
necessarily occupied---with degeneracies which depend on the way the
block is chosen and which are different from the non-interacting case.
We thus expect an evolution with increasing $U$ from an equidistant
spectrum at $U=0$ through some intermediate regime into a different
equidistant spectrum at $U/t\gg 1$. Such an evolution is indeed
observed, see in particular figure
\ref{Fig_Int_EntSpectrum_Middle}(b), where already at $U=10$ the
spectrum becomes again almost equidistant.

\begin{figure}
\begin{center}
\includegraphics*[width=0.99\columnwidth]{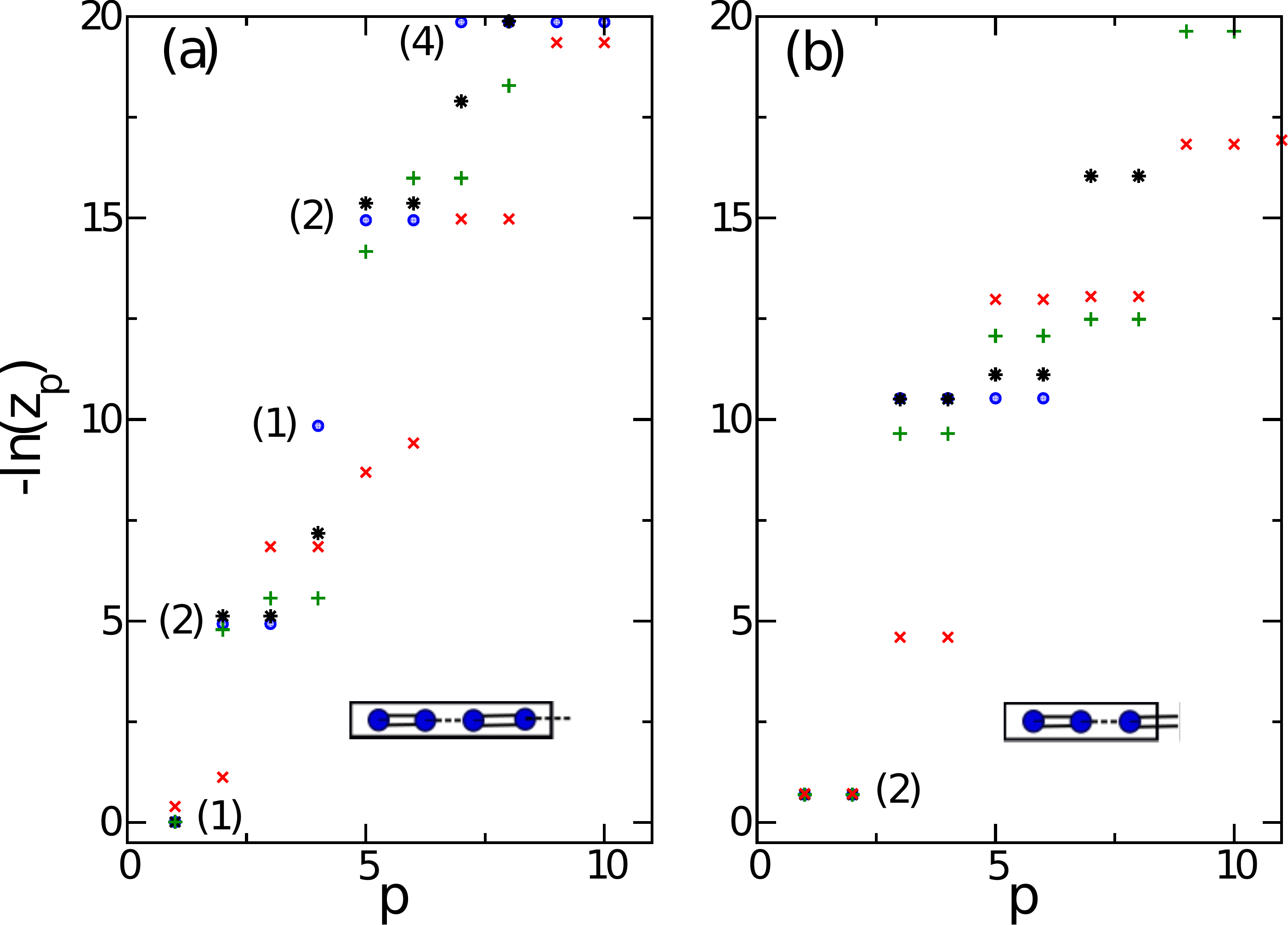}
\end{center}
\caption{Entanglement spectrum for a block at the end of a chain of
  length $N=20$ at $\delta=-0.5$ with (a) even block size $\ell=4$,
  and (b) odd block size $\ell=5$.  The blue circles are data for
  $U=0$, the black stars for $U =0.5$, the green pluses for $U=2$, and
  the red crosses for $U=10$.  The numbers in parenthesis give the
  degeneracy of the $U=0$ spectra.}
\label{Fig_Int_EntSpectrum_End1}
\end{figure}

Next we plot the entanglement spectrum for a block at the end of a
chain, a measure which is sensitive to the presence of localized edge
states.  First, we consider the evolution of the spectrum in what is
the topological trivial phase for $U=0$, see figure
\ref{Fig_Int_EntSpectrum_End1}, cutting (a) a weak bond, and (b) a
strong bond.
Again we find that the degeneracy of the lowest level is stable up to
intermediate interactions, $U=2$, while a new level structure with a
twofold degenerate ground state emerges in both cases for $U=10$.

\begin{figure}
\begin{center}
\includegraphics*[width=0.99\columnwidth]{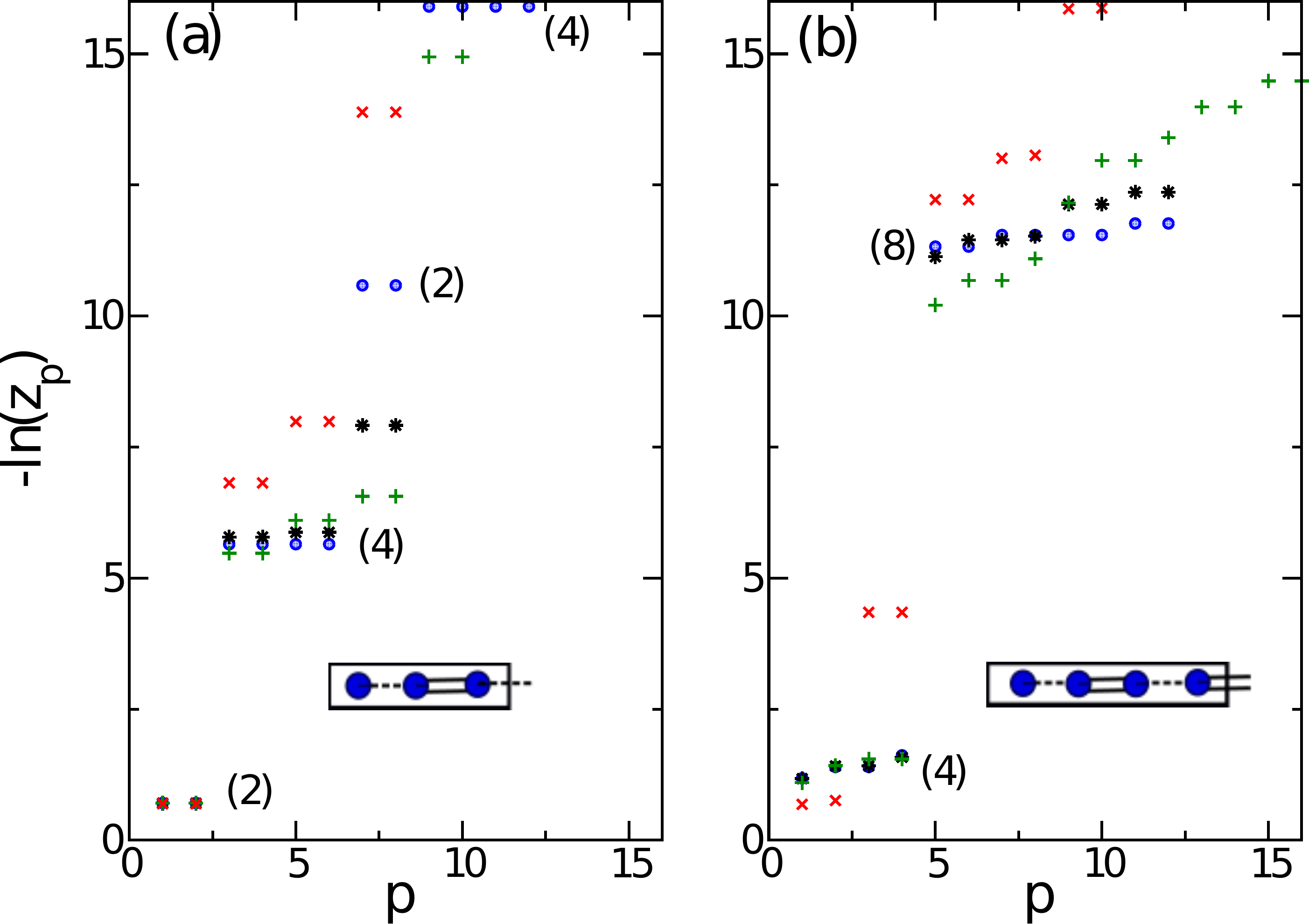}
\end{center}
\caption{Same as figure \ref{Fig_Int_EntSpectrum_End1} but for
  $\delta=0.5$ with (a) $\ell=5$, and (b) $\ell=4$.}
\label{Fig_Int_EntSpectrum_End2}
\end{figure}

For a block at the end of a chain in the SPT phase, shown in figure
\ref{Fig_Int_EntSpectrum_End2}, the picture is qualitatively similar.
The boundary susceptibility and entanglement spectra for $\delta<0$
and $\delta>0$ thus remain distinct from each other even if
nearest-neighbour interactions of intermediate strength are added. In
particular, we still find numerically a twofold (fourfold) degeneracy
of the lowest level for a block with $\ell$ odd ($\ell$ even) at the
end of a chain in the SPT phase, see figure
\ref{Fig_Int_EntSpectrum_End2}. For $U=0$ this is directly related to
an edge state which is shared between the two boundaries and which is
thus, for an infinite separation of the boundaries, only partially
filled.

\subsection{LCRG results for infinite chains}
\begin{figure}
\begin{center}
\includegraphics*[width=0.99\columnwidth]{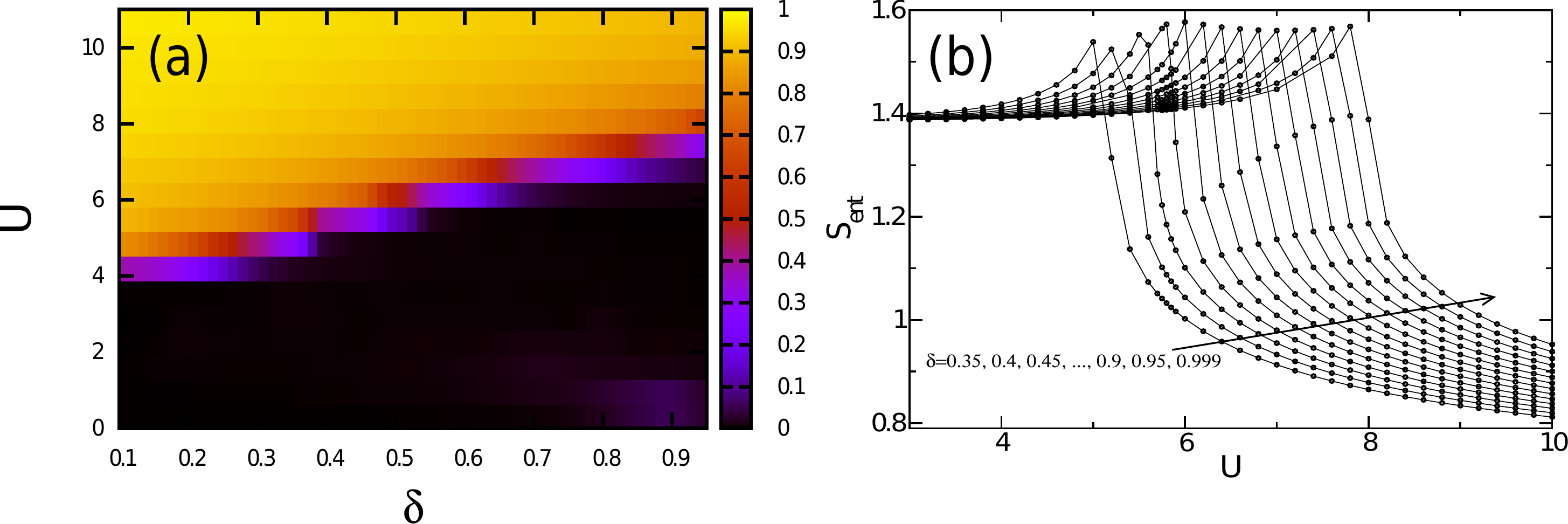}
\end{center}
\caption{(a) Degeneracy of the lowest level, measured by
  $3z_1-z_2-z_3-z_4$, and (b) entanglement entropy $S_{\rm {ent}}$ of
  the reduced density matrix for an effectively infinite block in an
  infinite chain obtained by LCRG.}
\label{Fig_LCRG1}
\end{figure}

The LCRG algorithm is a variant of the density-matrix renormalization
group \cite{WhiteDMRG,EnssSirker}. Originally proposed to study the
real time evolution after a quantum quench, it is also possible to
consider an imaginary time evolution starting from an initial product
state. In this case the initial state is projected onto the ground
state of the Hamiltonian used in the time evolution and ground state
properties of infinite chains can be studied. Here we will use this
algorithm to investigate the entanglement spectrum and entropy of an
essentially infinite block within the infinite chain as a function of
dimerization $\delta$ and interaction $U$. We choose our block in such
a way that two strong bonds are cut so that the largest eigenvalue of
the reduced density matrix for $U=0$ will be fourfold degenerate. In
figure \ref{Fig_Int_EntSpectrum_Middle} we have seen that this
fourfold degeneracy changes into a twofold degeneracy at large
interaction strengths. In order to find the point where this change
occurs, we order the eigenvalues of the reduced density matrix by
magnitude and plot LCRG results for $3z_1-z_2-z_3-z_4$ in figure
\ref{Fig_LCRG1}(a).  This quantity is zero as long as the largest
eigenvalue is fourfold degenerate while it should become non-zero in
the CDW phase. Indeed, a very sharp transition is observed numerically
with $U_c\approx 4$ at $\delta=0.1$ and $U_c\approx 8$ for $\delta\to
1$. Since the symmetries which protect the SPT phase are not violated
by the interaction, the excitation gap should close at the phase
transition.  To confirm a closing of the energy gap at the transition,
we show the numerically calculated entanglement entropy as a function
of $U$ for various dimerizations $\delta$ in figure
\ref{Fig_LCRG1}(b).
The entanglement entropy evolves in all cases shown from
$S_{\rm{ent}}\approx 2\ln 2$ at $U=0$ to $S_{\rm{ent}}=\ln 2$ for
$U\to\infty$. At $U$ values consistent with those where we find an
abrupt change of the degeneracy of the largest eigenvalue, see figure
\ref{Fig_LCRG1}(a), the entanglement entropy has a sharp maximum. Here
a comment about the numerical calculations is in order: We time evolve
the initial product state up to imaginary times times $\tau=50$. For
all $U$ values apart from the small region where $S_{\rm{ent}}$ is
peaked the projection converges and the values shown in figure
\ref{Fig_LCRG1}(b) are the finite fully converged entropies of the
gapped ground states. At the phase transition, however, we find
$S_{\rm{ent}}\sim\ln\tau$ consistent with a diverging entanglement
entropy for $\tau\to\infty$ as expected for a gapless state. In this
case the projection is not fully converged and the values for
$S_{\rm{ent}}$ shown are simply the values obtained at our maximal
simulation time $\tau=50$.

\section{Summary and Conclusions}
\label{Conclusions}
A system in a symmetry protected topological (SPT) phase is an
insulator with short-range entanglement. The scaling of the bulk
entanglement entropy with block size for a block in an infinite chain
as well as the bulk fidelity are therefore not any different from
those of a topologically trivial insulator. SPT phases, however, have
localized edge states so that the entanglement properties for a block
at the end of a chain with SPT order as well as the boundary fidelity
susceptibility are expected to deviate from those in a topological
trivial phase.

To investigate these boundary contributions we have considered a
well-known example: a chain of non-interacting spinless fermions with
a dimerized hopping in which an SPT phase is realized for $\delta>0$
while the phase for $\delta<0$ is topologically trivial. We have
confirmed that the bulk fidelity only depends on the absolute value of
the dimerization $\delta$, i.e., it cannot be used to distinguish
between the SPT and the trivial phase. For the boundary contribution
$\chi_B$, on the other hand, we have found that $\chi_B(\delta<0)<0$
while $\chi_B(\delta>0)>0$. Furthermore, the divergence of $\chi_B$ at
the critical point $\delta=0$ seems to follow a power law with
exponents which are different in the two phases.

For a block in an infinite chain we have tested a prediction from
massive field theory \cite{CardyCastro-Alvaredo} for the scaling of
the entanglement entropy with block size $\ell$ for the case $\ell
\lesssim\xi$ where $\xi$ is the correlation length. Apart from a
recent numerical study of spin models \cite{LeviCastro-Alvaredo} this
is the second independent numerical confirmation of this formula for a
different class of gapped models. Again, SPT and trivial phase show
exactly the same properties. This is different if a block at the end
of a semi-infinite chain is considered instead. In this case we have
shown that the scaling with block size in the trivial phase is still
apparently of the same functional form as for a block in an infinite
chain while the scaling is drastically altered in the SPT phase.  If
the boundary state is completely filled we find, in particular, a
maximum in $S_{\rm{ent}}(\ell)$ at a finite block size $\ell$. I.e.,
the saturation value $S_{\rm{ent}}(\ell\to\infty)$ is now approached
from above. We have explained this maximum as a direct consequence of
additional entanglement caused by the boundary state which is
exponentially localized on the scale of the localization length of
this state. In the entanglement spectrum this qualitatively different
behaviour in the SPT phase is reflected by a rather smooth spectrum
for $\ell\lesssim\xi$ while only for $\ell\gg\xi$ the equidistant
spectrum, obtained by corner transfer methods for a large block in an
infinite system, is recovered. In the latter limit the spectra in the
SPT and in the trivial phase for a block at the end of a semi-infinite
chain thus become identical again.

Finally, we have used exact diagonalization and Arnoldi algorithms to
investigate in how far the distinct boundary fidelity and entanglement
properties in the SPT phase survive if interactions are included. To
this end, we have considered the dimerized chain with a
nearest-neighbour repulsion, which, in the limit $U/t\to\infty$, will
drive the system into a charge-density-wave state. Our results show
that both the different sign of the boundary susceptibility in the SPT
as compared to the trivial phase as well as the exponentially
localized additional entanglement for a block at the end of a chain in
an SPT phase are stable features up to the phase transition into the
topological trivial CDW phase. Using a density-matrix renormalization
group algorithm for infinite chains we have, furthermore, shown that
the phase transition can be located precisely by finding the point
where the degeneracy of the largest eigenvalue of the reduced density
matrix changes.

For the future it would be of interest to develop a massive field
theory approach to explain the scaling of the entanglement entropy
found in this work for a block at the end of a semi-infinite chain
both in topological trivial as well as in SPT phases.

\ack We acknowledge support by the DFG via the collaborative research
center SFB/TR 49, the Graduate School of Excellence MAINZ, and the
Natural Sciences and Engineering Council (NSERC) of Canada. We are
grateful to the Regional Computing Center at the University of
Kaiserslautern, the AHRP, and Compute Canada for providing
computational resources and support.\\

\providecommand{\newblock}{}


\begin{thebibliography}{10}
\expandafter\ifx\csname url\endcsname\relax
  \def\url#1{{\tt #1}}\fi
\expandafter\ifx\csname urlprefix\endcsname\relax\def\urlprefix{URL }\fi
\providecommand{\eprint}[2][]{\url{#2}}

\bibitem{SSH}
Su W, Schrieffer J and Heeger A 1979 {\em Phys. Rev. Lett.\/} {\bf 42} 1698

\bibitem{SuSchriefferHeegerRMP}
Heeger A~J, Kivelson S, Schrieffer J~R and Su W~P 1988 {\em Rev. Mod. Phys.\/}
  {\bf 60} 781

\bibitem{RyuSchnyder}
Ryu S, Schnyder A~P, Furusaki A and Ludwig A~W~W 2010 {\em New Journal of
  Physics\/} {\bf 12} 065010

\bibitem{JackiwRebbi}
Jackiw R and Rebbi C 1976 {\em Phys. Rev. D\/} {\bf 13} 3398

\bibitem{RyuHatsugai02}
Ryu S and Hatsugai Y 2002 {\em Phys. Rev. Lett.\/} {\bf 89} 077002

\bibitem{Wen12}
Wen X~G 2012 {\em Phys. Rev. B\/} {\bf 85} 085103

\bibitem{Berry}
Berry M~V 1984 {\em Proc. R. Soc. A\/} {\bf 392} 45

\bibitem{Zak}
Zak J 1989 {\em Phys. Rev. Lett.\/} {\bf 62} 2747

\bibitem{RyuHatsugai06}
Ryu S and Hatsugai Y 2006 {\em Phys. Rev. B\/} {\bf 73} 245115

\bibitem{ViyuelaRivas}
Viyuela O, Rivas A and Martin-Delgado M\ A 2014 {\em Phys. Rev. Lett.\/} {\bf
  112} 130401

\bibitem{FidkowskiKitaev}
Fidkowski L and Kitaev A 2011 {\em Phys. Rev. B\/} {\bf 83} 075103

\bibitem{PollmannTurner}
Pollmann F, Turner A~M, Berg E and Oshikawa M 2010 {\em Phys. Rev. B\/} {\bf
  81} 064439

\bibitem{TurnerPollmann}
Turner A~M, Pollmann F and Berg E 2011 {\em Phys. Rev. B\/} {\bf 83} 075102

\bibitem{AbastoHamma}
Abasto D~F, Hamma A and Zanardi P 2008 {\em Phys. Rev. A\/} {\bf 78} 010301

\bibitem{ShuoGu}
Yang S, Gu S~J, Sun C~P and Lin H~Q 2008 {\em Phys. Rev. A\/} {\bf 78} 012304

\bibitem{Shin}
Shin B~C 1997 {\em Bull. Austral. Math. Soc.\/} {\bf 55} 249

\bibitem{AndersonOC}
Anderson P~W 1967 {\em Phys. Rev. Lett.\/} {\bf 18} 1049

\bibitem{ZanardiPaunkovic}
Zanardi P and Paunkovic N 2006 {\em Phys. Rev. E\/} {\bf 74} 031123

\bibitem{VenutiZanardi}
Venuti L~C and Zanardi P 2007 {\em Phys. Rev. Lett.\/} {\bf 99} 095701

\bibitem{RamsDamski}
Rams M~M and Damski B 2011 {\em Phys. Rev. Lett.\/} {\bf 106} 055701

\bibitem{SirkerFidelity}
Sirker J 2010 {\em Phys. Rev. Lett.\/} {\bf 105} 117203

\bibitem{SirkerHerzog}
Sirker J, Herzog A, Ole\'s A~M and Horsch P 2008 {\em Phys. Rev. Lett.\/} {\bf
  101} 157204

\bibitem{CalabreseCardy}
Calabrese P and Cardy J 2004 {\em J. Stat. Mech.\/}  P06002

\bibitem{SirkerEntanglement}
Sirker J 2012 {\em J. Stat. Mech. P12012\/}

\bibitem{JinKorepin}
Jin B~Q and Korepin V~E 2004 {\em J. Stat. Phys.\/} {\bf 116} 79

\bibitem{AffleckLudwig}
Affleck I and Ludwig A~W~W 1991 {\em Phys. Rev. Lett.\/} {\bf 67} 161

\bibitem{LaflorencieSoerensen}
Laflorencie N, S\o{}rensen E~S, Chang M~S and Affleck I 2006 {\em Phys. Rev.
  Lett.\/} {\bf 96} 100603

\bibitem{CalabreseCampostrini}
Calabrese P, Campostrini M, Essler F and Nienhuis B 2010 {\em Phys. Rev.
  Lett.\/} {\bf 104} 095701

\bibitem{CardyCalabrese10}
Cardy J and Calabrese P 2010 {\em J. Stat. Mech.\/}  P04023

\bibitem{FagottiCalabrese}
Fagotti M and Calabrese P 2011 {\em J. Stat. Mech.\/}  P01017

\bibitem{DalmonteErcolessi1}
Dalmonte M, Ercolessi E and Taddia L 2011 {\em Phys. Rev. B\/} {\bf 84} 085110

\bibitem{DalmonteErcolessi2}
Dalmonte M, Ercolessi E and Taddia L 2012 {\em Phys. Rev. B\/} {\bf 85} 165112

\bibitem{CardyCastro-Alvaredo}
Cardy J, Castro-Alvaredo O and Doyon B 2008 {\em J. Stat. Phys.\/} {\bf 130}
  129

\bibitem{LeviCastro-Alvaredo}
Levi E, Castro-Alvaredo O~A and Doyon B 2013 {\em Phys. Rev. B\/} {\bf 88}
  094439

\bibitem{ChungPeschel}
Chung M~C and Peschel I 2001 {\em Phys. Rev. B\/} {\bf 64} 064412

\bibitem{Peschel02}
Peschel I 2002 {\em J. Phys. A: Math. Gen.\/} {\bf 36} L205

\bibitem{KluemperScheeren}
Kl\"umper A, Martinez J~R~R, Scheeren C and Shiroishi M 2001 {\em J. Stat.
  Phys.\/} {\bf 102} 937

\bibitem{SirkerKluemperEPL}
Sirker J and Kl\"umper A 2002 {\em Europhys. Lett.\/} {\bf 60} 262

\bibitem{PeschelEisler}
Peschel I and Eisler V 2009 {\em J. Phys. A\/} {\bf 42} 504003

\bibitem{ManmanaEssin}
Manmana S~R, Essin A~M, Noack R~M and Gurarie V 2012 {\em Phys. Rev. B\/} {\bf
  86} 205119

\bibitem{EnssSirker}
Enss T and Sirker J 2012 {\em New J. Phys.\/} {\bf 14} 023008

\bibitem{AlbaHaque}
Alba V, Haque M and L\"auchli A~M 2012 {\em Phys. Rev. Lett.\/} {\bf 108}
  227201

\bibitem{WhiteDMRG}
White S~R 1992 {\em Phys. Rev. Lett.\/} {\bf 69} 2863

\end{thebibliography}

\end{document}